\pdfoutput=1
\documentclass[a4paper, 11pt]{article}
\usepackage{jheppub}
\usepackage{graphicx}
\usepackage{overpic}
\usepackage{amsmath}
\usepackage{sidecap}
\usepackage{verbatim} 
\usepackage{url}
\usepackage{setspace}
\usepackage{xspace}
\usepackage{afterpage}

\usepackage{color}

\newcommand{\pT}{$p_{T}$\xspace} 
 
\newcommand{\conv}{\texttt{Conv2D}\xspace}
\newcommand{\convtranspose}{\texttt{Conv2DTranspose}\xspace}
\newcommand{\filtermask}{\texttt{FilterMask}\xspace}

\newcommand{\antikt}{anti-$k_{t}$\xspace}
\newcommand{\kt}{$k_{t}$\xspace}

\title{Deep Learning as a Parton Shower}

\author{J.~W.~Monk}
\affiliation{Niels Bohr Institute, \\ University of Copenhagen, Denmark}
\emailAdd{jmonk@cern.ch}
\arxivnumber{1807.03685}

\abstract{We make the connection between certain deep learning architectures and the renormalisation group explicit in the context of QCD by using a deep learning network to construct a toy parton shower model. The model aims to describe proton-proton collisions at the Large Hadron Collider. A convolutional autoencoder learns a set of kernels that efficiently encode the behaviour of fully showered QCD collision events. The network is structured recursively so as to ensure self-similarity, and the number of trained network parameters is low. Randomness is introduced via a novel custom masking layer, which also preserves existing parton splittings by using layer-skipping connections. By applying  a shower merging procedure, the network can be evaluated on unshowered events produced by a matrix element calculation. The trained network behaves as a parton shower that qualitatively reproduces jet-based observables.}

\begin{document}

\maketitle

\section{Introduction}\label{sec:introduction}

The renormalisation group provides a set of rules that describe how a system evolves under a re-scaling transformation.  This is expressed in parton shower models by the repeated evaluation of a splitting kernel over an ordered hierarchy of scales, which results in the self-similarity that is a characteristic of renormalisable models.  We have previously exploited this behaviour by using wavelet decomposition \cite{Monk:2014uza} to extract features from radiation patterns in proton-proton  collision events at both small and large angles.

A single layer in a convolutional neural network (CNN) is very similar to a single level wavelet decomposition.  Indeed, with appropriate network parameters, the CNN \emph{is} a wavelet decomposition.  Since the wavelet basis can be used to reveal the angular evolution of a parton shower, this raises the intriguing possibility that a CNN could be structured in such a way that it encodes, and behaves as, a parton shower model.   That the hierarchical structure of deep learning architectures is formally connected to the behaviour of the renormalisation group is an area of active interest, see e.g. \cite{2014arXiv1410.3831M, 2013arXiv1301.3124B, 2017arXiv170511023O}; in this paper we will make the connection between these ideas obvious by constructing a toy parton shower model using a deep learning neural network whose design has been inspired by wavelet decomposition.

\begin{figure}[!t]
\begin{center}
\begin{overpic}[width=0.6\textwidth]{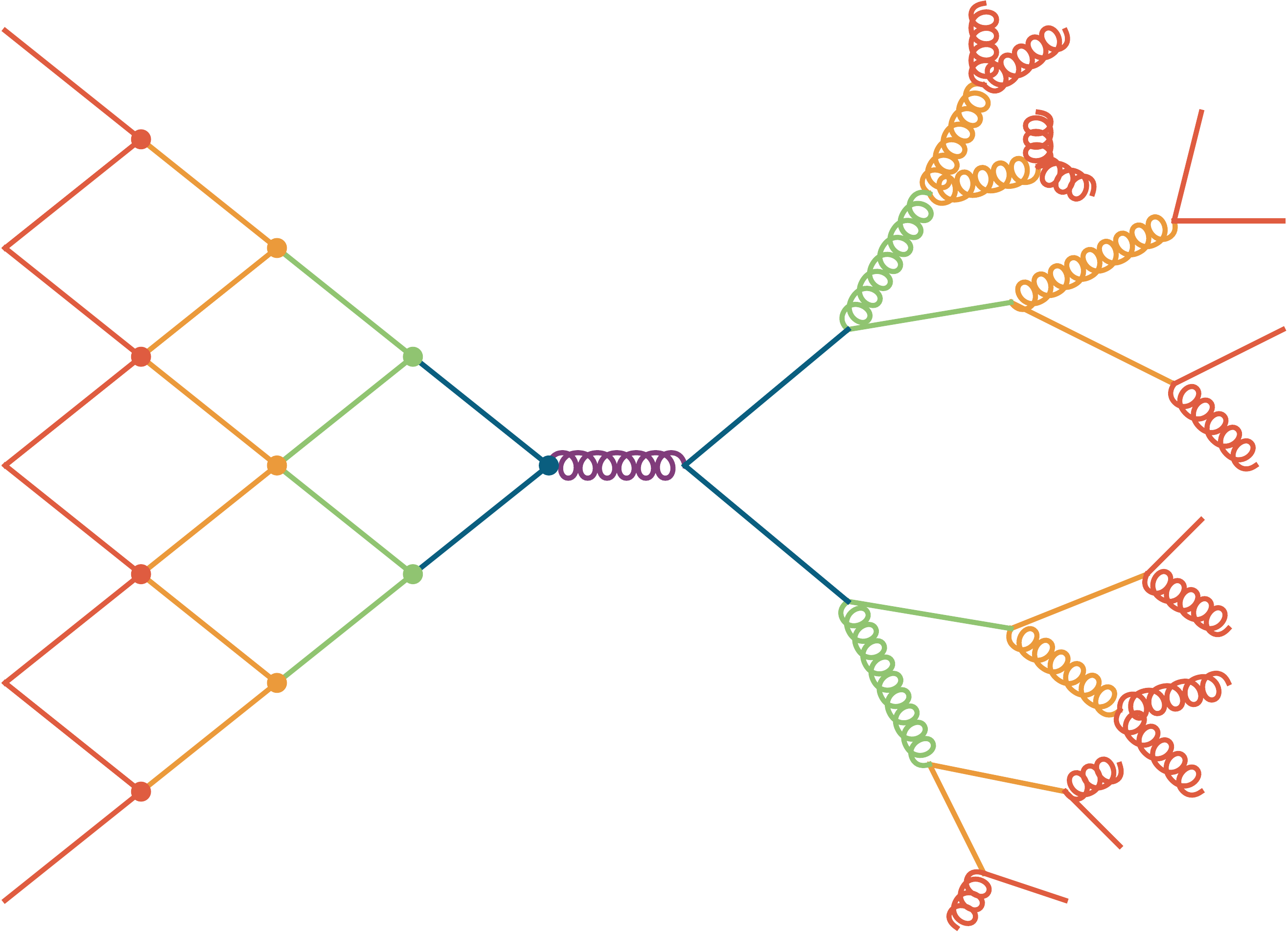}
\put(-31, 44){input pattern}
\put(98, 28){output pattern}
\end{overpic}
\caption[Autoencoding CNN as a shower]{Comparison between the structure of a convolutional autoencoder and a parton shower.  A pattern is input on the left and the value at each network node is determined by the weighted sum of the connected nodes.}\label{fig:CNNShower}
\end{center}
\end{figure}

An autoencoding neural network takes an input of high dimensionality, compresses it to a bottleneck of a small number of network nodes, then reinflates the compressed values to recover the input data as the target.  In so-doing, the behaviour of the input data is encoded in the network parameters.  The compression stage of a convolutional autoencoder uses a series of convolutional layers interspersed with pooling layers to repeatedly reduce the dimensionality of the input.  Having compressed the data at the bottleneck, the re-inflation half of the autoencoder again uses (different) convolutional layers interspersed with up-scaling.  Figure \ref{fig:CNNShower} shows the general form of a convolutional autoencoder, where the use of gluon lines is intended to make explicit the similarity between the re-inflation stage and an iterative parton shower.

The scaling behaviour of a parton shower means that the (de-) convolutional kernels used in each layer of the autoencoding CNN should be related to those used in all the other layers.  While effects like colour coherence or the divergence of the strong coupling at low energy mean that QCD is not exactly scale invariant, parton showers are used in an energy regime where scale invariance is a good approximation.  In practice, this means that the same convolutional kernels can be used in each layer, which ensures self-similarity over the different angular scales that the network layers represent.  Despite the re-use of the same kernels at each scale, the running of the strong coupling can be approximated by evolving the relative contributions of the different available kernels.  The re-use of the same kernel in multiple layers also means there are a relatively small number of independent network parameters, even if the network is deep. This multi-scale coarse-graining approach means that behaviour that the network learns at one angular scale is applied to all angular scales down to the cut-off.

Other results have also noted that the self-similarity of parton showers can be encoded in a recursive neural network.  For example, the JUNIPR model described in \cite{Andreassen:2018apy} uses a recursive network based on hierarchical $1\rightarrow 2$ particle splittings to learn probability distributions for particle emissions within jets.  Conversely, image-based network designs such as \cite{deOliveira:2017pjk} have also been used to generate jet images directly.  However, as far as we are aware, our model is the first that combines both recursion with an image-based approach that can generate entire events and that can be merged with a fixed order matrix element calculation.  A more detailed discussion of the similarities and differences between the three approaches will be presented in  section \ref{sec:conclusion}.

Building a deep learning network that can approximate the behaviour of QCD is a useful exercise for several reasons: such a network can help provide insights into \emph{why} neural networks (sometimes!) work so well for analysis tasks; the network can extract features and observables directly from data, which can be used to confront existing shower models; the evolution of the network parameters with depth in the network can provide some insight into the structure of showers;  the trained autoencoder will not fit data that is different from that it is trained on, hence could be used to identify signal that differs from the QCD backgound; and the toy model trained directly on data can provide a useful comparison to existing methods for tuning Monte Carlo models.

\section{Design of the Autoencoding CNN}\label{sec:design}

\subsection{Network structure}
\begin{figure}[!ht]
\includegraphics[width=\textwidth]{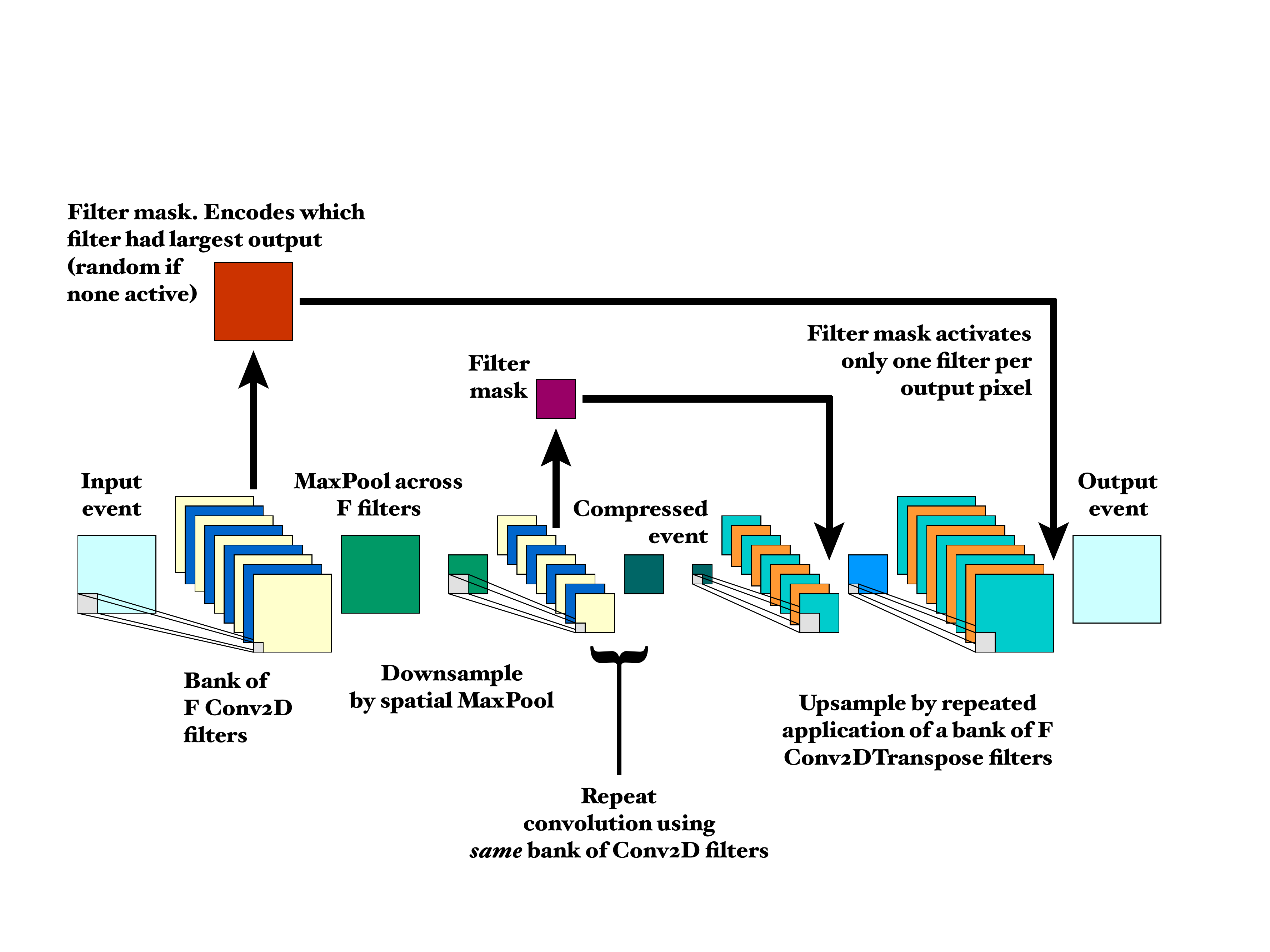}
\caption[Autoencoder design]{Overview of the autoencoding CNN.  An input event image enters the network on the left and is repeatedly processed by a bank of $F$ \conv filters.  Having been completely compressed, the event is reinflated using \convtranspose layers together with a special \filtermask layer.}\label{fig:convnet}
\end{figure}

The layout of the autoencoding CNN is shown in Figure \ref{fig:convnet}. A set of $F$ 2-dimensional convolutional layers (\conv) is used in the compression half of the autoencoder, and $F$ transposed 2-dimensional convolution layers (\convtranspose) are used in the reinflation half of the autoencoder.  The \conv layers are defined with kernels of size $k\times k$ and pad the input so that the output from the layer has the same size as the input.  The \convtranspose layers are also defined with kernels of size $k\times k$, but use a stride size of $k$ together with padding of the input to ensure that the output from the layer has dimensions $kN\times kN$, where the dimensionality of the input to the layer is $N\times N$.  Event data is provided to the autoencoding CNN model in the form of pixel arrays that represent the emissions of energy from each proton-proton collision. 

The input image is passed through the $F$ \conv layers, which results in a stack of $F$ output images, each of which is the same size as the original input image.  A max-pooling layer is used across these $F$ images so that each pixel site output by the max-pooling layer is the maximum value of the corresponding pixel sites in the $F$ input images.  The output of the max-pooling across the $F$ convolutions is thus a single image the same size as the initial input image.  This single image is then downsampled by a further spatial max-pooling layer.  The spatial max-pooling uses a pool size of $k$, meaning that an initial image of size $kN\times kN$ is downsampled to a $N\times N$ image.   The combined effect of the filter max-pooling followed by spatial max-pooling is to combine a $k\times k$ region of pixels into a single pixel, using the filter that best matches the shape of the input for that region.

This $N\times N$ image is once again passed to the \emph{same} set of $F$ \conv layers, followed by the max-pooling layers, to further reduce the dimensionality of the image.  The sequence of convolution followed by max-pooling is repeated until the output image size is $k\times k$.  Note that if such a $k\times k$ image were again to be passed through the \conv and max-pooling layers, the result would be a single pixel.

The fully compressed $k\times k$ image is then passed to the set of $F$ \convtranspose layers, which results in a stack of $F$ $k^{2}\times k^{2}$ images that can be written as a single tensor $T_{ijk}$, where index $i$ is in the range $\left\{0\mathrel{{.}\,{.}}F\right\}$ and $j$ and $k$ are both in the range $\left\{0\mathrel{{.}\,{.}} k^{2}\right\}$.  This stack of up-scaled images is converted to a single image by using a custom layer that we have named \filtermask.  The \filtermask layer, $M_{ijk}$, uses the corresponding $k^{2}\times k^{2} \rightarrow k\times k$ downsampling in the compression stage of the autoencoder to decide which of the pixels in $T_{ijk}$ should be used.  \filtermask is a tensor with the same $F\times k^{2}\times k^{2}$ shape as $T_{ijk}$, but in which each pixel is either zero or one, as described in equation \ref{eqn:filtermask}  

\begin{equation}
M_{ijk} = \begin{cases}
1 & \forall m, C_{ijk} \geq C_{mjk}\\
0 & \mathrm{otherwise}
\end{cases}\label{eqn:filtermask}
\end{equation}
where $C_{ijk}$ is the output of the corresponding compression filters on the compression stage of the network.  Pixels are one if the corresponding pixel in the \conv output image is the maximum in that stack of pixels.  All other pixels in the \filtermask are zero.  The output stack of images from the \convtranspose is multiplied by the mask $M_{ijk}$, so that at most one pixel in the stack is non-zero.  This set of masked images is then converted to a single image, $S_{jk}$, by summing all of the pixels in a stack as given in equation \ref{eqn:mergestack}

\begin{equation}
S_{jk} = \sum_{i=0}^{F}M_{ijk}T_{ijk}\label{eqn:mergestack}
\end{equation}

The \filtermask transfers information about which \conv filter was active in the compression stage to the reinflation stage, meaning that the \convtranspose filter kernels are dependent on the upstream \conv kernels.  This also means that there is a mechanism by which a splitting present in the input event can be preserved through the network, while admitting a degree of randomness.  The \filtermask keeps a counter of the number of times each \conv filter was active during training, and converts these rates into a set of probabilities that have unit sum.  In the case that all \conv filters produce identical (therefore zero) output in the same pixel, the \filtermask randomly picks a single filter to activate according to its recorded probability.  This means that if a single active pixel is passed into the CNN, it will be propagated to the $k\times k$ bottleneck as a single pixel but then reinflated using randomly activated \convtranspose upscaling filters.  It is this feature of the \filtermask \ - preserving input splittings when they exist, while producing random splittings when none are present - that allows the autoencoding CNN to behave as a parton shower.

Having inflated the $k\times k$ image to $k^{2}\times k^{2}$ and applied the \filtermask derived from the $k^{2}\times k^{2} \rightarrow k\times k$ compression stage, the output $k^{2}\times k^{2}$ image is once again passed to the stack of \convtranspose filters, which output a stack of $k^{3}\times k^{3}$ images.  A \filtermask is once again used to merge the stack, but this time the mask is taken from the $k^{3}\times k^{3}\rightarrow k^{2}\times k^{2}$ compression stage.  Different \filtermask layers, with different learned filter probabilities, are thus used at the different compression and reinflation levels.  This allows the filter activation probabilities to evolve with angular scale.

The process of applying the stack of \convtranspose filters, followed by merging with the corresponding \filtermask from the compression stage, is repeated until the image is the same size as the original input image.

The CNN is implemented in python using Keras \cite{keras} with the TensorFlow \cite{2016arXiv160304467A} backend and is evaluated using a pair of Nvidia 1080 Ti graphics processing units (GPUs).  The code is available from \cite{APEMEN}.

\subsection{Loss Function}

The loss function of a neural network is the objective that should be minimised in order to best describe the input data.  A common simple loss function for a convolutional autoencoder is to take the mean squared error (MSE) between the input and output image, summing over all the pixels.  Minimising the MSE means that the output of the network is as similar as possible to the input.  

However, there are two problems with such a loss function.  MSE is very susceptible to aliasing effects in which an output emission is in a neighbouring pixel to a similar input emission.  The MSE penalises the network equally whether it produces an emission near to a target emission or very far away.  The second problem is that the input event images are sparse; there are 4096 pixels in a $64\times64$ grid, but a single event may contain only $\mathcal{O}\left(10-100\right)$ emissions.  Using a naive MSE loss means that the CNN will mainly learn about empty pixels, and will be biased towards producing no output activity.

These two problems are solved by modifying the naive MSE.\footnote{A theoretically nicer alternative might be to run a jet-finding algorithm, such as the \kt algorithm, on the output and the target, and compare infra-red safe jets.  Using different jets with both large and small radius parameters would provide sensitivity to both wide- and small-angle emissions, down to a carefully chosen cut off.  However, implementing the \kt algorithm on a GPU within TensorFlow is distinctly non-trivial, and offloading the image data from the GPU to the CPU for evaluation is computationally prohibitive.}  Both the input target and the CNN output are blurred by using a set of truncated Gaussian kernels.  A MSE-type loss is calculated for each Gaussian kernel, and a weighted sum of the losses is performed as in equation \ref{eqn:LossSum}

\begin{align}\nonumber
T^{i}_{\gamma \delta} &= \sum\limits_{\mu \lambda} T_{\alpha \beta} G^{i}_{\alpha-\gamma, \beta-\delta}  \\ \nonumber
O^{i}_{\gamma \delta} &= \sum\limits_{\mu \lambda} O_{\alpha \beta} G^{i}_{\alpha-\gamma, \beta-\delta}  \\
\mathcal{L} &= \frac{\sum\limits_{i}w_{i}M\left(T^{i}_{\gamma\delta}, O^{i}_{\gamma\delta}\right)}{\sum\limits_{i}w_{i}} \label{eqn:LossSum}
\end{align}

where $T_{\alpha\beta}$ is the target event image that is input to the CNN and $O_{\alpha\beta}$ is the output image of the CNN.   $G^{i}_{\gamma\delta}$ are a set of truncated Gaussian kernels and $T^{i}_{\gamma\delta}$ and $O^{i}_{\gamma\delta}$ are versions of the input and output, respectively, that have been blurred by $G^{i}_{\gamma\delta}$.  The loss function, $\mathcal{L}$, is the weighted sum over an MSE-like function, $M\left(T^{i}_{\gamma\delta}, O^{i}_{\gamma\delta}\right)$ using weights $w_{i}$.  The truncated Gaussian kernels are given by equation \ref{eqn:GaussianKernels}

\begin{equation}
G^{1} = \begin{bmatrix} 
1\\ 
\end{bmatrix}, G^{2}=\begin{bmatrix}
0.25 & 0.25 \\
0.25 & 0.25 \\
\end{bmatrix}, G^{3}=\begin{bmatrix}
0.0947 & 0.118 & 0.0947 \\
0.118 & 0.148 & 0.118 \\
0.0947 & 0.118 & 0.0947 \\
\end{bmatrix}\label{eqn:GaussianKernels}
\end{equation}

The MSE-like function has two contributions, one from pixels in which there is activity in the target image, and one from pixels in which there is no target activity.  The function $M\left(T^{i}, O^{i}\right)$ is given in equation \ref{eqn:MSE}

\begin{equation}
M\left(T^{i}, O^{i}\right) = \frac{\sum\limits_{\alpha,\beta}\left(T^{i}_{\alpha\beta} - O^{i}_{\alpha\beta}\right)^{2}\mathcal{M}_{\alpha\beta}}{\sum\limits_{\alpha,\beta}\mathcal{M}_{\alpha\beta}} + \frac{\left(\sum\limits_{\alpha,\beta}O^{i}_{\alpha\beta}\left(1-\mathcal{M}_{\alpha\beta}\right)\right)^{2}}{\sum\limits_{\alpha,\beta}\left(1-\mathcal{M}_{\alpha\beta}\right)}
\label{eqn:MSE}
\end{equation}
where $\mathcal{M}_{\alpha\beta}$ is a mask image whose pixels have value 1 if the corresponding pixel in $T^{i}_{\alpha\beta}$ is non-zero, and are zero otherwise.  The denominators  in equation \ref{eqn:MSE} account for the fact that events contain different numbers of target pixels with non-zero values.  Without this normalisation by the number of active pixels there could be a bias towards fitting events that contain more target emissions.  The normalisation ensures that the active and empty regions are given equal weight in the loss, while  the overall effect of equation \ref{eqn:MSE} is to treat all pixels that do not have any target activity in them as a single pixel.

The weights, $w_{i}$, give some control over the degree to which the loss penalises the network for producing activity at a large angle to a target emission.  Increasing weight $w_{3}$, which applies to the $G^{3}$ Gaussian kernel, causes the loss function to reduce the penalty for producing emissions at a wide angle to the target emission.  Nevertheless, regardless of the weights chosen, the loss is always minimised by producing output in exactly the same pixel as the target.   Apart from $w_{i}$, the custom loss does not introduce any additional model hyper parameters and the behaviour of the loss is determined by its functional form, which is chosen to stabilise the loss during training relative to a standard MSE.  

\subsection{Regularisation of Network Kernels}

During training of the network, it may be possible for the convolutional kernel weights to diverge or become infinitesimal.  In order to prevent this, dropout layers that randomly mask the output of the convolutional layers would typically be used to regularise the kernel weights.  However, dropout cannot be used here because the max-pooling across convolutional filters is non-linear.   Dropout works well when the network approximates a linear summation of neurons because it allows many subsets of the available network structures to be explored.  However, in this model the convolutional kernels interact with each other, so it is not possible to drop a single network node without radically altering the network behaviour.

In lieu of dropout, a regularisation penalty is applied to prevent the learned convolutional kernel values from diverging.  The kernels are similar to shower splittings, and so a regularisation term is added that penalises kernels that deviate from energy conservation.  The kernel penalty term for each \conv filter with kernel weights $\vec{C}$ is given by equation \ref{eqn:energyConservation}

\begin{equation}
\mathcal{R}\left(\vec{C}\right)=\lambda \times \left(1-\sum\limits_{i=1}^{k}C_{i}\right)^{2}
\label{eqn:energyConservation}
\end{equation}
where $\lambda$ is a multiplier that controls the strength of the regularisation and the summation is over all of the kernel weights in the \conv layer.  The penalty term for each kernel is calculated using equation \ref{eqn:energyConservation} and added to the total loss for the model state during training. 

The aim of the regularisation is to prevent the model reaching a state in which only a single filter is active during training so that interactions between filters can produce the desired complex behaviour.  Since the \filtermask layer couples the activation rates of the \conv and \convtranspose filters, it is sufficient to apply regularisation only to the \conv layers in order to prevent such convergence.  In addition, if regularisation is applied to both the \conv and \convtranspose filters then the space of possible filter configurations is greatly reduced and the model cannot achieve the desired complexity.  Therefore regularisation is only applied to the \conv filters and is not applied to the \convtranspose filters.

\subsection{Model Parameters}\label{sec:params}

Having provided the general network structure, two sets of hyper-parameters define two concrete implementations of the CNN model.  Model $k_{2}$ uses a kernel size of $k=2$, and model $k_{3}$ uses a kernel size of $k=3$.  The max-pooling and up-scaling used by the model requires that the input pixel array dimensions, $N\times N$, must obey the rule $N=k^{n}$, where $n$ is an integer. Model $k_{2}$ is defined using input pixel arrays of size $64\times 64$, and model $k_{3}$ uses inputs of size $81\times 81$.  The kernel and input array size together define the number of levels of convolution that the model performs; model $k_{2}$ is a narrower but deeper model with more levels of decomposition, while model $k_{3}$ uses a wider kernel but fewer levels of decomposition.  The model hyper-parameters and other details are given in table \ref{tab:hyper}

\begin{table}
\centering
\begin{tabular}{c | c | c }
parameter & model $k_{2}$ & model $k_{3}$ \\
\hline
Kernel size, $k$ & 2 & 3 \\
Input image size, $N$ & 64 & 81 \\
Size of filter bank, $F$ & 9 & 7 \\
Levels of decomposition & 5 & 3 \\
Regularisation, $\lambda$ & 500 & 300 \\
Learning rate & $\mathrm{5\times 10^{-5}}$ & $\mathrm{1\times 10^{-5}}$ \\
Loss weight $w_{1}$ & 5 & 4 \\
Loss weight $w_{2}$ & 2 & 2 \\
Loss weight $w_{3}$ & 1 & 1 \\
Total number of trained weights & 72 & 126 \\
\end{tabular}\caption[Model hyper-parameters]{Model hyper-parameters}\label{tab:hyper}
\end{table}

The choice  of the size of the filter bank, $F$, is initially inspired by the desire for rotational invariance in the $k_{2}$ model.  With $k=2$ there are four possible rotations of the kernel, plus a parity transformation, meaning that eight filters can cover all possible transformations of one kernel.  One additional filter is added in order to allow for a non-splitting.    For the larger $k_{3}$ kernel, it is found that training a corresponding filter bank size of $F=17$ is prohibitive, so the number of filters is reduced to $F=7$.  

The interaction between the filter kernels means that the behaviour of the model changes rather non-linearly with the size of the filter bank.  If $F$ is too small the model lacks the complexity to describe parton production, but if $F$ is too large then the model complexity grows so much that training becomes very difficult.  It is serendipitous that our initial choice of a filter bank size capable of providing exact rotational invariance in model $k_{2}$ is also about the right size to provide a viable model complexity.  There remains sufficient flexibility in the model that the smaller filter bank in model $k_{3}$  is still capable of encoding the shower.  Future improvements to the training procedure, either in the loss function or the optimisation algorithm, may permit a larger filter bank.

The other model hyper-parameters are decided by training models with different sets of parameter values using a small sample of the training events and inspecting the evolution of the training and validation losses.  The size of the filter bank has by far the largest effect on the model, with the other parameters being of secondary importance.  The regularisation $\lambda$ can to some extent be used to control the complexity of the trained model.  Lower $\lambda$ values produce models that are easier to train but that are less complex and consequently produce less chaotic emissions, and generally fewer emissions overall.  Values of lambda in the range 100-1000 have been explored.  The learning rate must be small enough that the model does not jump over the minimum in the loss function.  The learning rates of table \ref{tab:hyper} are found to be sufficiently small; smaller values would also work, but would prolong the training time.  The values of the loss weights, $w_{i}$, only enter the loss as a ratio because they are normalised by the sum of the weights.  The choice of the weights values depends on the resolution of the images used with the model.  Smaller pixels reduce the blur radius of the Gaussian filter, so the weight $w_{1}$ is slightly reduced for mode $k_{3}$, which has a slightly higher image resolution compared to mode $k_{2}$.

\section{Training and Evaluating the Models}

\subsection{Monte Carlo Event Samples and Selection}

Simulated samples of proton-proton events are needed in order to train and to test the CNN models.  Sherpa 2.2.4 \cite{Gleisberg:2008ta, Schumann:2007mg, Krauss:2001iv, Gleisberg:2008fv, Hoeche:2009rj} is used to generate a sample of 8.5 million QCD proton-proton collision events with up to four outgoing parton legs in the matrix element calculation.  The default Sherpa tune is used, with the NNPDF 3.0 PDF set \cite{Ball:2014uwa} and a shower merging scale of 20~GeV.  The beam energy is 6.5~TeV per proton.  Hadronisation and multi-parton interactions (MPI) are turned off because they have different scaling characteristics to a pure parton shower model, and would therefore require a more complicated deep learning  model than is studied here.\footnote{Hadronisation is in any case a small-angle effect, and should not have a large contribution given that the events are converted into pixel arrays with finite pixel sizes.  MPI will require further study to implement as part of a deep learning network, but given that MPI effects can be extracted by using a simple threshold in wavelet-space, it seems hopeful that much of the effect of MPI could be removed from training data by applying a  threshold on network layer activity.}   The shower turn off scale is left at the Sherpa default value of 3~GeV.  Sherpa's internal event selector is used to ensure that at least two R=0.4 jets with \pT greater than 25~GeV are produced by the matrix element calculation.

A subsequent event selection is made on the post-showering final-state particles that requires at least two R=0.4 \antikt jets \cite{Cacciari:2008gp} with \pT greater than 40~GeV and rapidity satisfying $\left|y\right| < \left(\pi -0.4\right)$ in each event.  The \antikt jet algorithm is run via the FastJet \cite{Cacciari:2011ma} library.  This criterion selects approximately 0.5 million events from the initial 8.5 million generated events.

The CNN model requires pixel arrays as inputs.  For this first implementation, a square pixel array using square pixels is used in order to avoid any unforeseen complications that might arise from using a image dimensionality that is not symmetric in rapidity-azimuth ($y - \phi$).  However, there is in principle no reason why this network design could not be extended to a larger rapidity range by using more pixels in the rapidity dimension.  Each of the selected Sherpa events is converted to $N\times N$ pixel-array images in rapidity-azimuth by identifying the pixel in the $y - \phi$ plane into which each particle is emitted and adding the particle \pT to the pixel value.  Pixels have a value of zero if no particles are emitted into them.  The pixel array covers the rapidity range $-\pi \le y < \pi$ and the azimuthal range $0 \le \phi < 2\pi$.  Each pixel array is normalised by dividing by the total sum of pixel values in the array and multiplying by $N\times N$ so that the average pixel value is one.  The normalisation has no distorting effect on the event or the model, whose kernels learn about \emph{ratios} between pixel values (which are unchanged by normalisation).  However, normalisation is helpful in avoiding numerical issues that could otherwise potentially bias the model towards events with more activity.  After normalisation, the arrays contain no information about the overall energy in the event, and the model can only learn about the shape of the radiation patterns.

Pixel arrays of size $64\times 64$ and $81\times 81$ are produced for use with models $k_{2}$ and $k_{3}$, respectively.  Larger pixel arrays could be used, but due to the requirement that $N=k^{n}$, they would need to be a minimum of $128\times 128$ and $243\times 243$, which makes training the model considerably more difficult.  Studies of the angular separation between individual particles produced by Sherpa's shower show that pixel sizes around $0.1\times 0.1$ have sufficient resolution to capture the vast majority of the details of the radiation patterns.  Pixel arrays are therefore limited to $64\times 64$ and $81\times 81$.

In order to test the trained CNN shower models, unshowered partons produced by a matrix element calculation are needed.  An additional sample of 8.5 million matrix element (ME) Sherpa events generated without any parton shower is used for this purpose.  These events have the same generator-level process and selection of two R=0.4 jets with \pT greater than 25~GeV, but lack any post-shower selection.  The full sample of 8.5 million ME events is converted to $N\times N$ pixel arrays, but is not normalised.  

A shower merging scheme for the CNN shower models is introduced in section \ref{sec:merging}.  As a test of this scheme, two further Sherpa samples of 8.5 million events - one with showering, one without - are produced with an alternative merging scale of 40~GeV (compared to 20~GeV for the nominal samples).  These alternative samples are not used for training and are only used to check that the results of the CNN merging scheme are not dependent on the merging scale.

\subsection{Training on Showered Events}

The fully showered event pixel arrays are divided into a training sample, which contains 90\% of the 0.5 million images, and a validation sample, which contains the remaining 10\%.  

\begin{figure}[!t]
\begin{overpic}[width=0.5\textwidth]{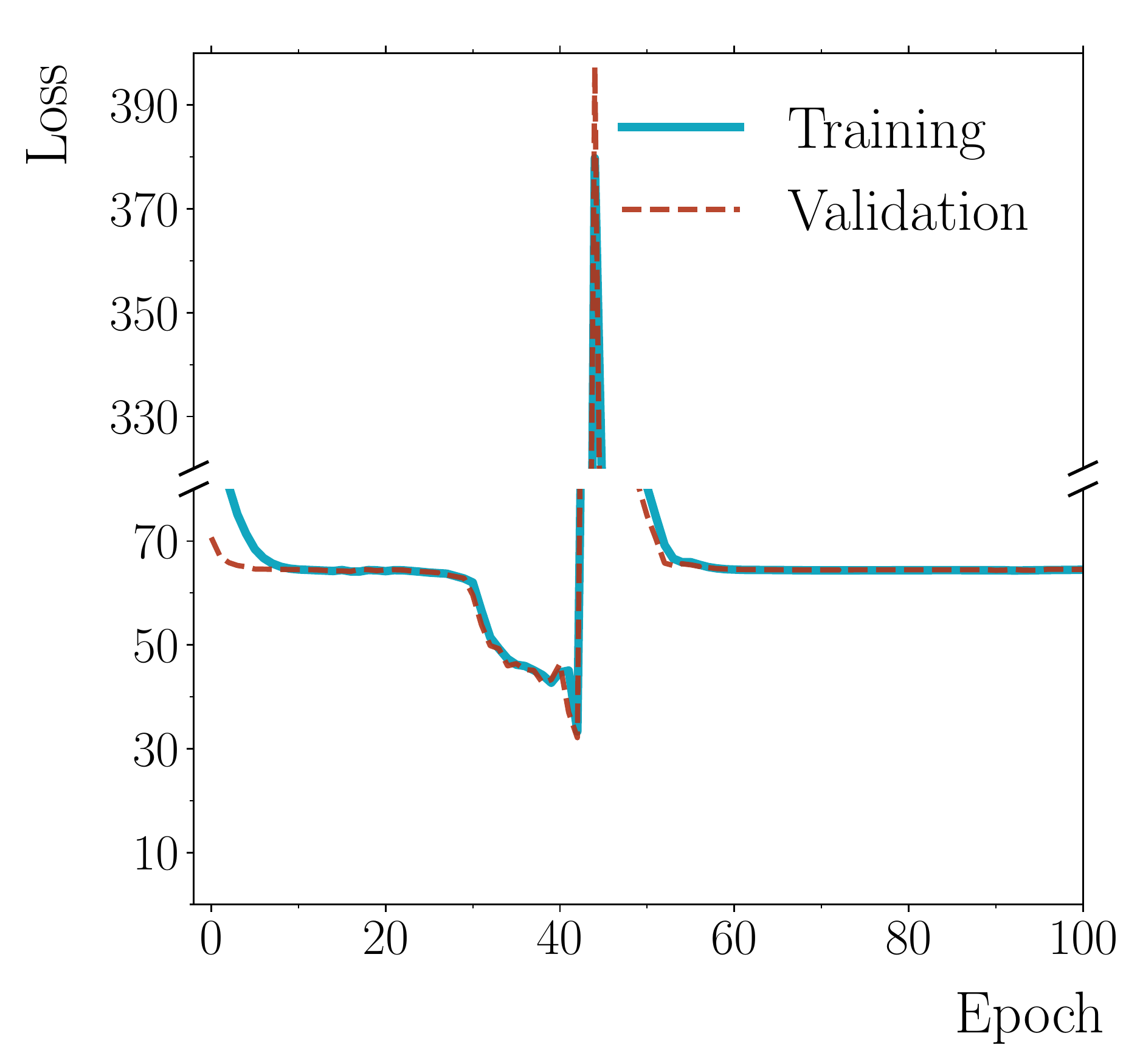}
\put(23, 80){a)}
\end{overpic}
\begin{overpic}[width=0.5\textwidth]{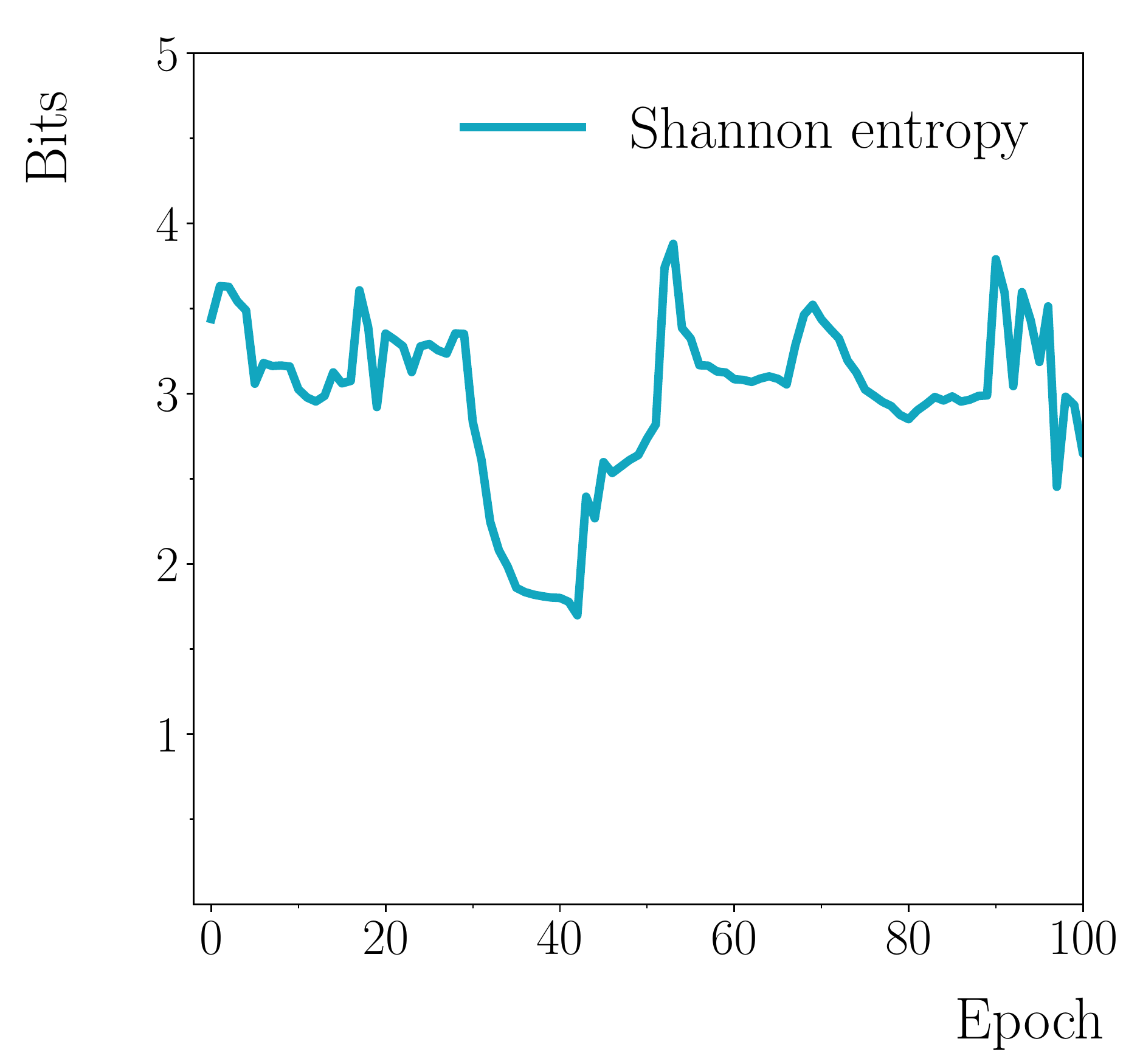}
\put(23, 80){b)}
\end{overpic}
\caption[Training]{Evolution of the model during training.  a), left, shows how the loss function evaluated on the training and validation event samples changes during training, while b), right, shows the evolution of the Shannon entropy of the \filtermask layer.}\label{fig:trainingloss}
\end{figure}

Models are trained for several hundred epochs using the learning rates given in Table \ref{tab:hyper}.  The Nadam optimiser \cite{2016dozat, 2016arXiv160904747R} is used and the network weights are initialised with random values from the Glorot Normal distribution \cite{GlorotAISTATS2010}.  An example of the evolution of the training and validation loss for model $k_{2}$ is shown in Figure \ref{fig:trainingloss}\,a).  During training, the model rests for some number of epochs in a moderate loss state, before falling into a lower loss state.  After spending a small number of epochs in the low loss state, the model then undergoes a rapid increase in both training and validation loss and reaches a high loss state, before falling once again to a (different) moderate loss state.  This evolution of both training and validation loss is shown in Figure \ref{fig:trainingloss}.  Note the lack of divergence between the training and validation loss, indicating that the rapid increase is not due to overfitting the training data.  The lack of divergence is seen regardless of the choice of model hyper-parameters (section \ref{sec:params}).  This lack of over-fitting is probably due to the high dimensionality of the input data together with the relatively high statistical power of the input and the rather small number of learned parameters.  The lack of divergence is also a good indicator that the training set of events is (more than) sufficient to reach the maximum potential performance of the trained model.

The probabilities stored in the \filtermask layers are the rates with which each filter is active, and the Shannon entropy\footnote{The Shannon entropy, $H$, of a single \filtermask layer is given by $H = \frac{\sum\limits_{i=0}^{F}p_{i}\ln{p_{i}}}{\ln{F}}$, where $p_{i}$ are the probabilities stored in the layer and $F$ is the size of the filter bank. } of those probabilities  is a measure of the complexity of the model state.  If the Shannon entropy is high, the filters are all active with similar rates, whereas if the entropy is low then a small number of the filters dominate the description of the collision events.  Figure \ref{fig:trainingloss}\,b) shows the evolution of the Shannon entropy of the model during the same training period as Figure \ref{fig:trainingloss}\,a).  The entropy is normalised so that each \filtermask layer encodes at most one bit of information, and for $k=2$ there are five \filtermask layers, so there is a maximum of five bits stored in the model.  At the start of training, the model is in a high entropy state because the filters are active with quasi-random probabilities.  When, during training, the loss falls to a minima, the entropy also declines to moderate values, showing that the model is in a more ordered state.  When the loss subsequently rises rapidly, the model undergoes a transition to a higher entropy state again.  When trained for a very large number of epochs (not shown here), the entropy of the model gradually evolves to a very low entropy state, while the loss remains on a plateau.  This long-term decline in entropy accompanied by a near-constant loss indicates that there are a large number of model states that are equally good at describing the collision events, and the model tends towards a state in which the model behaviour is dominated by a small number of the filters in the filter bank.  This behaviour of the model complexity could in future potentially be used in more effective network training algorithms.

Since the loss reaches a plateau after many training epochs, while the model complexity continues to reduce, the optimal model configuration is selected as the epoch that corresponds to the minimum loss prior to the rebound onto the plateau.  In the example of Figure \ref{fig:trainingloss} this corresponds to around the forty second epoch.  A callback function is used within Keras to save the corresponding best model state.

\subsection{Comparison With Class IV Cellular Automata}

Parton showers and the CNN implemented here are similar in both conception and behaviour to Cellular Automata (CA).  Cellular Automata evolve a system from an initial state using a set of rules that describe how the current state should change under a discrete step.  Parton showers employ a set of splitting rules to evolve the state of the shower between scales.  Similarly, the CNN uses rules described by the \conv or \convtranspose layers to step between angular scales.  The change in the CNN during training shares some interesting aspects with the change in behaviour of CA as they move through their available ``rule space.''

Cellular automata have been divided into four classifications \cite{WOLFRAM19841, 1990PhyD4577L, 2013arXiv1304.1242M}:

\begin{itemize}
\item Class I: The initial state evolves to a fixed pattern, and is not interesting for the present study
\item Class II: The evolution from the initial state is dominated by well-ordered periodic structures that are largely independent of the initial state.  
\item Class III: The system evolves chaotically and produces random patterns that are independent of the initial state.
\item Class IV: The system evolves to produce complex states that are neither completely chaotic, nor completely ordered.  The evolution is dependent on the initial state, and the rules interact in a non-linear way to produce complex behaviour.
\end{itemize}

One key indicator of the difference between classes II, III and IV is the entropy of the CA site activity.  Class II typically produces low entropy states, while Class III produces high entropy states.  As the rules of the CA are altered, it can undergo a (potentially rapid) phase transition between the classes.  

Prior to training,\footnote{The CNN should always be trained for one epoch on a small number of events to ensure that the probabilities in the \filtermask layers have been calculated correctly.  If the network is not trained at all, the probabilities will violate unitarity. As long as  training is sufficiently short, the network kernels will remain random.}  the CNN behaves very much like a Class III Cellular Automata.  The network weights are initialised to random values and produce a large number of random emissions, uncorrelated with the input to the network.  

After over-training\footnote{Over-training here means the model continues training long after the loss function has been minimised so that the Shannon entropy declines.} for many hundreds of epochs, the network behaves like a Class II Cellular Automata.  The Shannon entropy of the \filtermask layers declines, indicating that the model is highly-ordered and a small number of kernels dominate the evolution of the input through the network.

\begin{figure}[!ht]
\begin{overpic}[width=0.5\textwidth]{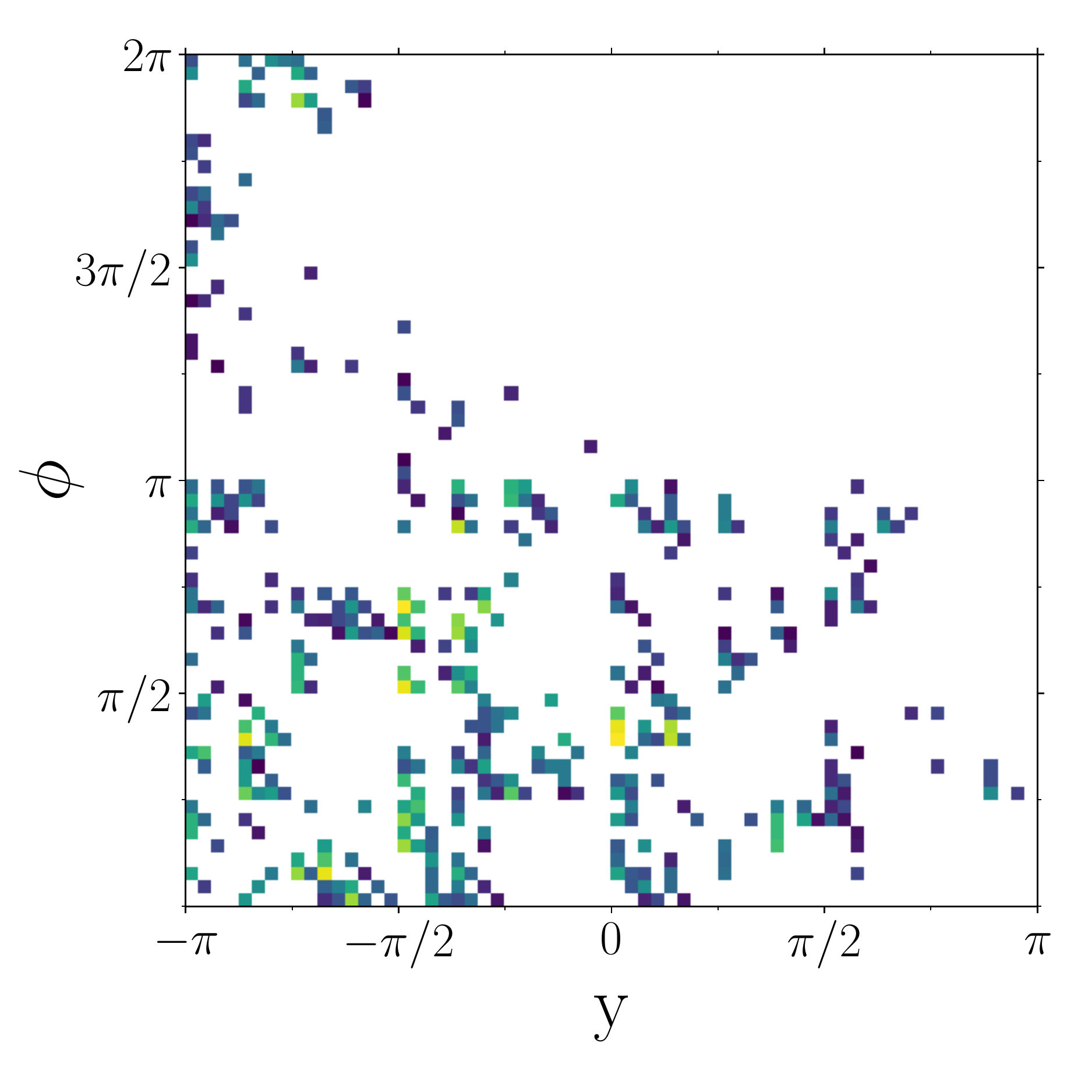}
\put(80,83){a)}
\end{overpic}
\begin{overpic}[width=0.5\textwidth]{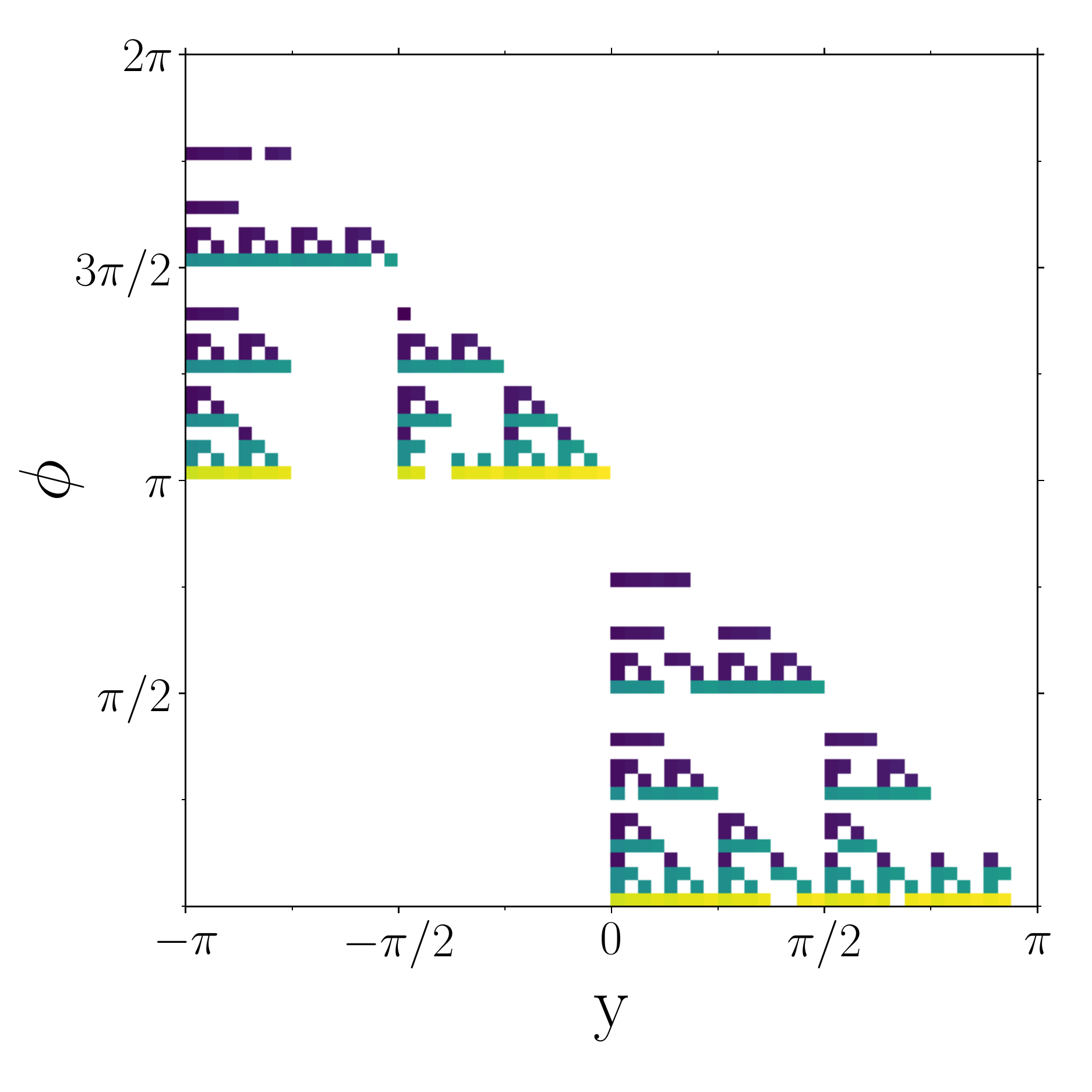}
\put(80,83){b)}
\end{overpic}
\caption[Over- and under-trained CNNs]{Example emission patterns produced by an untrained (chaotic) CNN (a) and an over-trained (periodic) CNN (b). The input to the CNN was the same in both cases.}\label{fig:cellularautomata}
\end{figure}

The evolution of the CNN from chaotic to highly ordered behaviour is illustrated by some radiation patterns from model $k_{2}$ in Figure \ref{fig:cellularautomata}.  Figure \ref{fig:cellularautomata}\,a) is the output of the untrained model $k_{2}$ when two random partons are used to seed the CNN.  The output in Figure \ref{fig:cellularautomata}\,a) is chaotic, there is no pattern and the output is uncorrelated with the input.  Figure \ref{fig:cellularautomata}\,b) shows the output from the same model $k_{2}$ using the same input when the model is over-trained by several hundred epochs.  The model output has become sparse, with well-ordered structures that are repeated over different angular scales.

As the rules of the CA are updated so that a transition from class II to class III is made, there can - depending on the specifics of the CA - be a Goldilocks region in which the CA is class IV and is capable of describing complex phenomena.  Similarly, as the CNN kernels are updated during training, it becomes capable of describing the complex behaviour of  a parton shower close to the transition from chaos to periodicity.  The goal for training a CNN capable of behaving as a parton shower should thus be to maximise the time spent exploring the kernel parameters in the transition region.

\subsection{Merging CNN with Matrix Element Calculations}\label{sec:merging}

The trained network is evaluated on the pixel arrays produced from Sherpa matrix element events that have not previously been showered.  The effect of the \filtermask layer in this case is to randomly activate filters according to the rate with which they were active during training.  The output of the evaluation on un-showered events is an approximation to fully showered events.  The output of the CNN is a pixel array of the same size as the input.  These output arrays are converted back into lists of particle-level collision events by creating a single particle for each pixel that has a \pT value above 100~MeV.  The particle \pT is the same as the pixel \pT, and the particle is emitted into a random location within the pixel.

Care must be taken to merge parton showers with perturbative matrix element calculations in order to prevent the double-counting of emissions from both the shower and the ME into the same region of phase space.  The Sherpa matrix elements used here assume a $k_{T}$ ordered parton shower, where $k_{T}$ is the transverse momentum of an emitted parton relative to the emitter.  Although the CNN does not currently explicitly define an ordering parameter, it most closely resembles an angular ordering.  This mis-match between the ordering used in the ME and the CNN implies that a veto should be applied to the CNN to prevent emission of shower partons into phase-space regions that should be covered by the ME \cite{Hamilton:2009ne}.  This shower merging is implemented as a new layer in the CNN that is added after each \convtranspose level.  The shower merging is only used during evaluation of the CNN on pixel arrays produced from matrix element events.  The shower merging is {\bf{not}} used during training on fully showered events.  

Each application of the \convtranspose filters corresponds to the generation of potential new parton emissions, which is why the shower merging veto is applied after each new application of the \filtermask and \convtranspose layers.  The shower veto scale evolves with depth in the CNN.  The network bottleneck in the middle of the autoencoder corresponds to the widest angle emissions, and is analogous to the shower starting scale.  The shower veto scale at the bottleneck is therefore determined from that provided by Sherpa's matrix element, $Q_{0}$, in this case $Q_{0}=$~20~GeV, in combination with the emission angle to which the bottleneck corresponds.  The emission angle, $\Delta\phi$, at the bottleneck is approximately given by    $\Delta\phi\simeq\pi/k$, and the shower veto scale, $Q$, at the bottleneck is therefore approximated by equation \ref{eqn:emissionAngle} 

\begin{equation}
Q\left(\Delta\phi\right)=Q_{0}/\left(1-\cos{\Delta\phi}\right)\label{eqn:emissionAngle}  
\end{equation}
Each de-convolutional level within the network corresponds to a different emission angle, and for each step in network level away from the bottleneck, the emission angle is divided by a factor of $k$.  The shower veto scale, $Q$, in a given network layer is thus determined from the running emission angle together with equation \ref{eqn:emissionAngle}  Thus the later layers of the re-inflation stage, which correspond to smaller angles, use a larger veto, which is appropriate as it allows collinear emissions more easily than wide angle emissions.

At each level of the re-inflation, the veto procedure is applied as follows:

\begin{itemize}
\item The output of the merged \convtranspose bank is sub-divided into windows the same size as the $k\times k$ convolutional kernel.
\item Any pixel below the veto scale for that convolutional level is left unchanged.
\item The number of pixels, $N_{S}$, in each window that are above the veto scale is calculated.
\item The number of pixels, $N_{ME}$, in each window of the corresponding layer on the compression side of the CNN is calculated.
\item If $N_{ME} \ge N_{S}$ then all pixels in the output window are accepted because no new emissions above the merging scale have been added in that region.
\item If $N_{ME} < N_{S}$ then the pixels above the merging scale in that window are replaced by those in the corresponding window of the image from the compression-side of the CNN.  These pixels correspond to the state of the shower prior to splitting.
\item Replacement hard pixels are adjusted to account for the emission of any soft pixels below the merging scale within the same window.  
\end{itemize}

This merging procedure can be carried out entirely via matrix manipulation operations, and is implemented as a Keras layer and executed on the GPU.  

\section{Jet Distributions Predicted by the CNN}

The Rivet analysis framework \cite{Buckley:2010ar} is used to compare events showered with Sherpa to events showered with the CNN, as well as ME-level events that have not been showered.   A sample of eight million events produced with Herwig 7.1.4 \cite{Marchesini:1983bm,Marchesini:1987cf,Gieseke:2003rz,Bahr:2008pv, ThePEG} is also used as an example of an angular ordered shower for comparison with Sherpa's $k_{T}$ ordered shower.  For these events, the default Herwig tune is used with a beam energy of 6500~GeV per proton.  The QCD $2\rightarrow 2$ process is used and, for a direct comparison with the Sherpa samples, both hadronisation and MPI are turned off.

Analyses are performed using two different jet algorithms from the FastJet package: the \antikt algorithm with a radius parameter of 0.4, and the \kt algorithm \cite{Ellis:1993tq} with a radius of 0.6.  Jets must have a  rapidity that satisfies $\left|y\right|<2.5$ and transverse momentum that satisfies \pT$> 40$~GeV.

\begin{figure}[!ht]
\begin{center}
\includegraphics[width=0.42\textwidth]{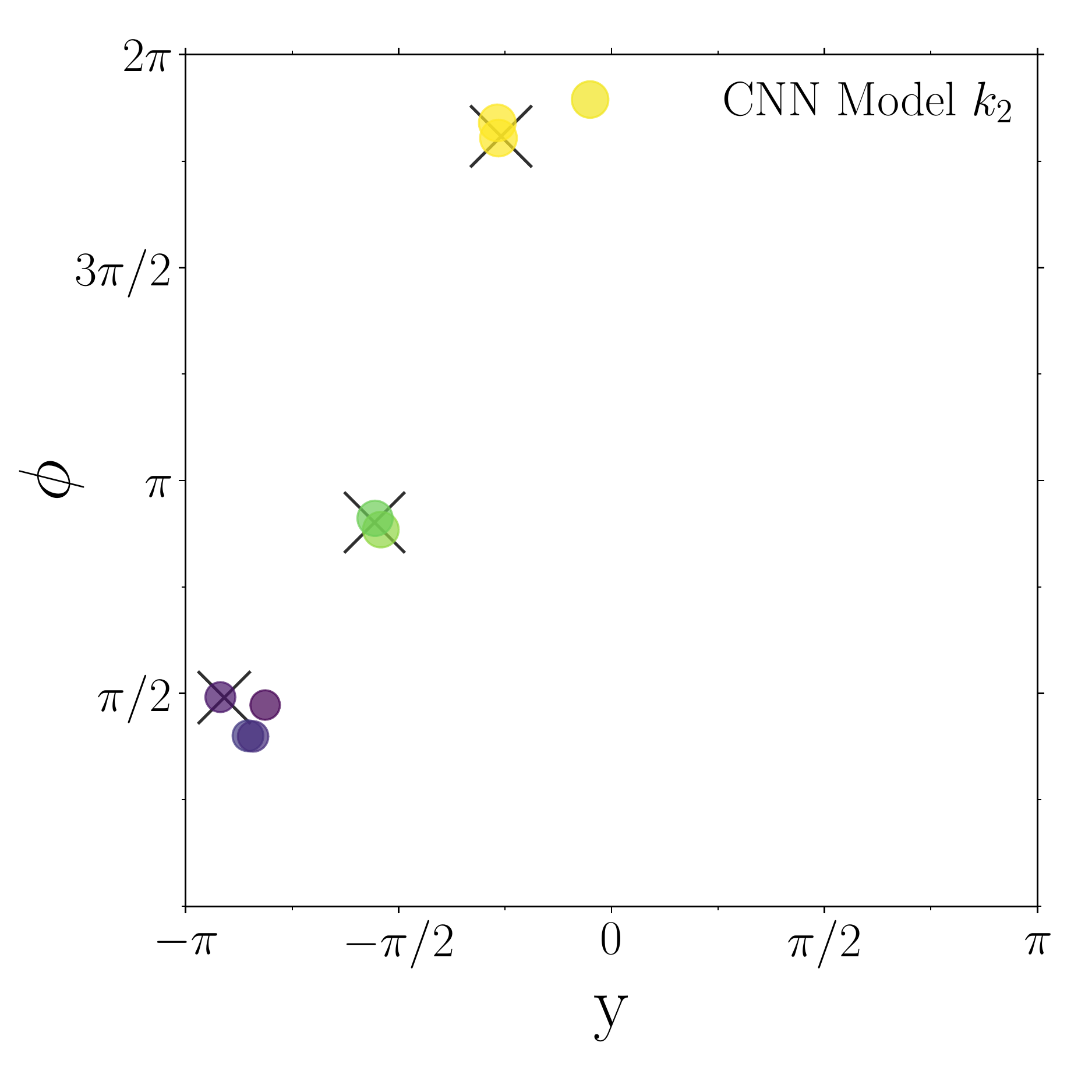}
\includegraphics[width=0.42\textwidth]{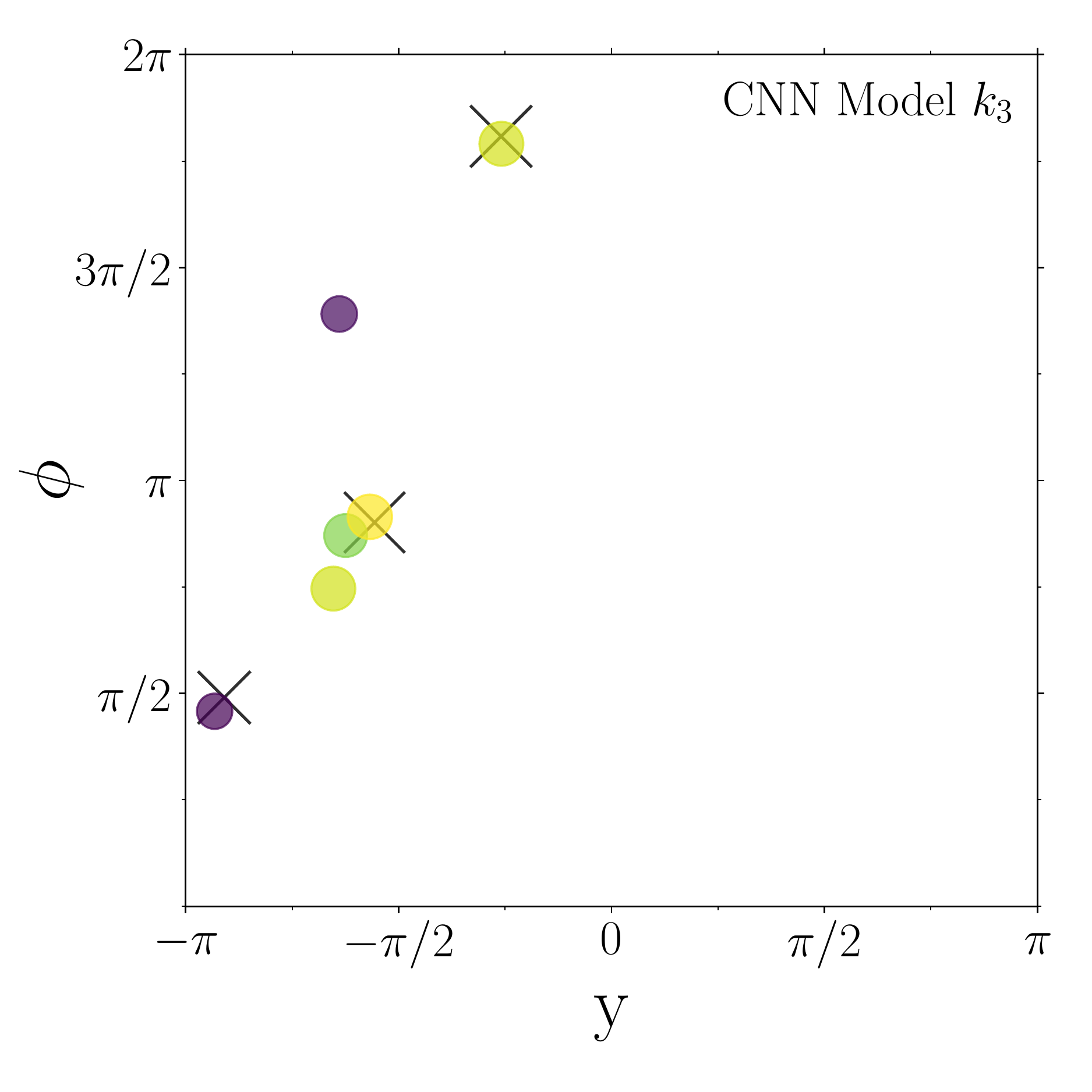}\\
\includegraphics[width=0.42\textwidth]{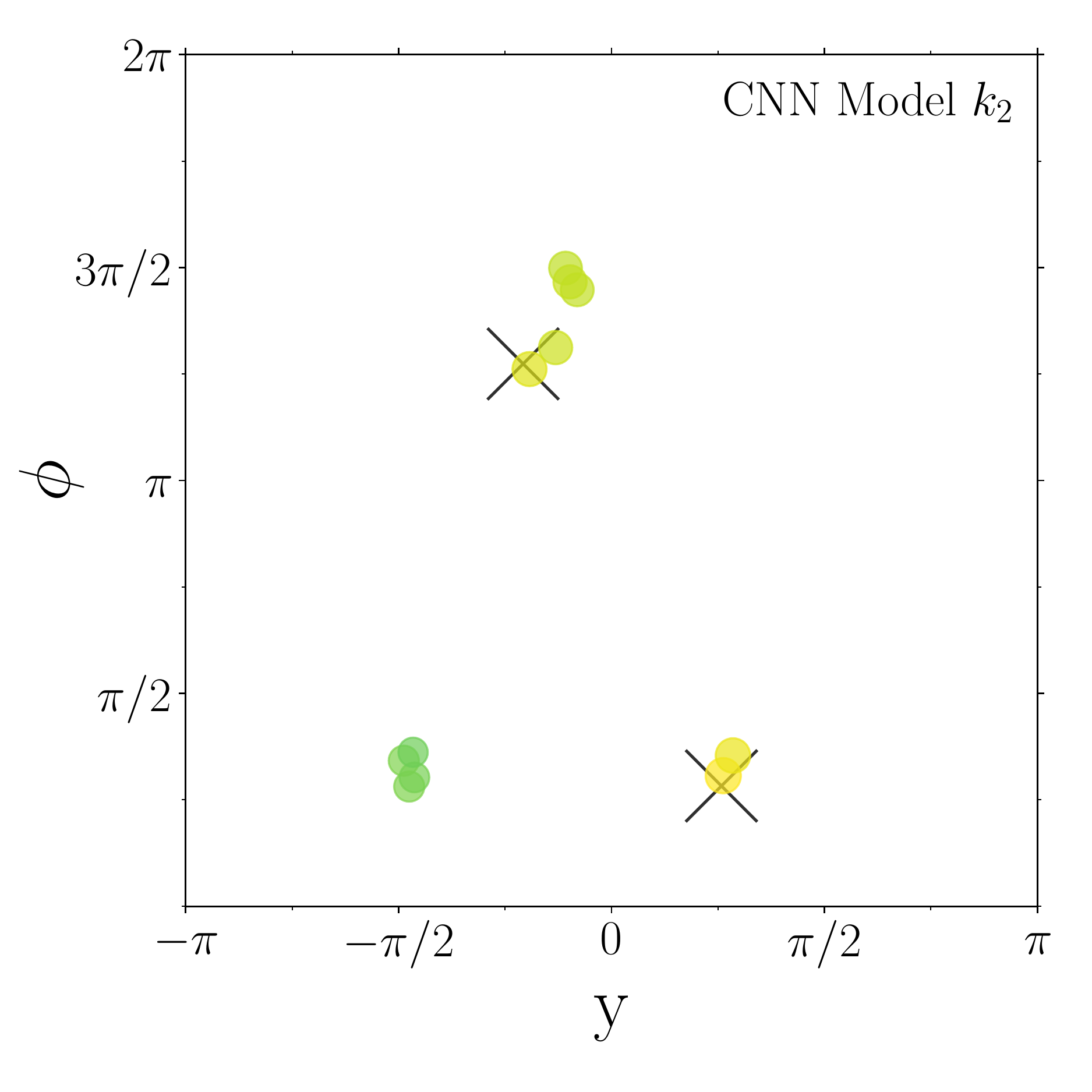}
\includegraphics[width=0.42\textwidth]{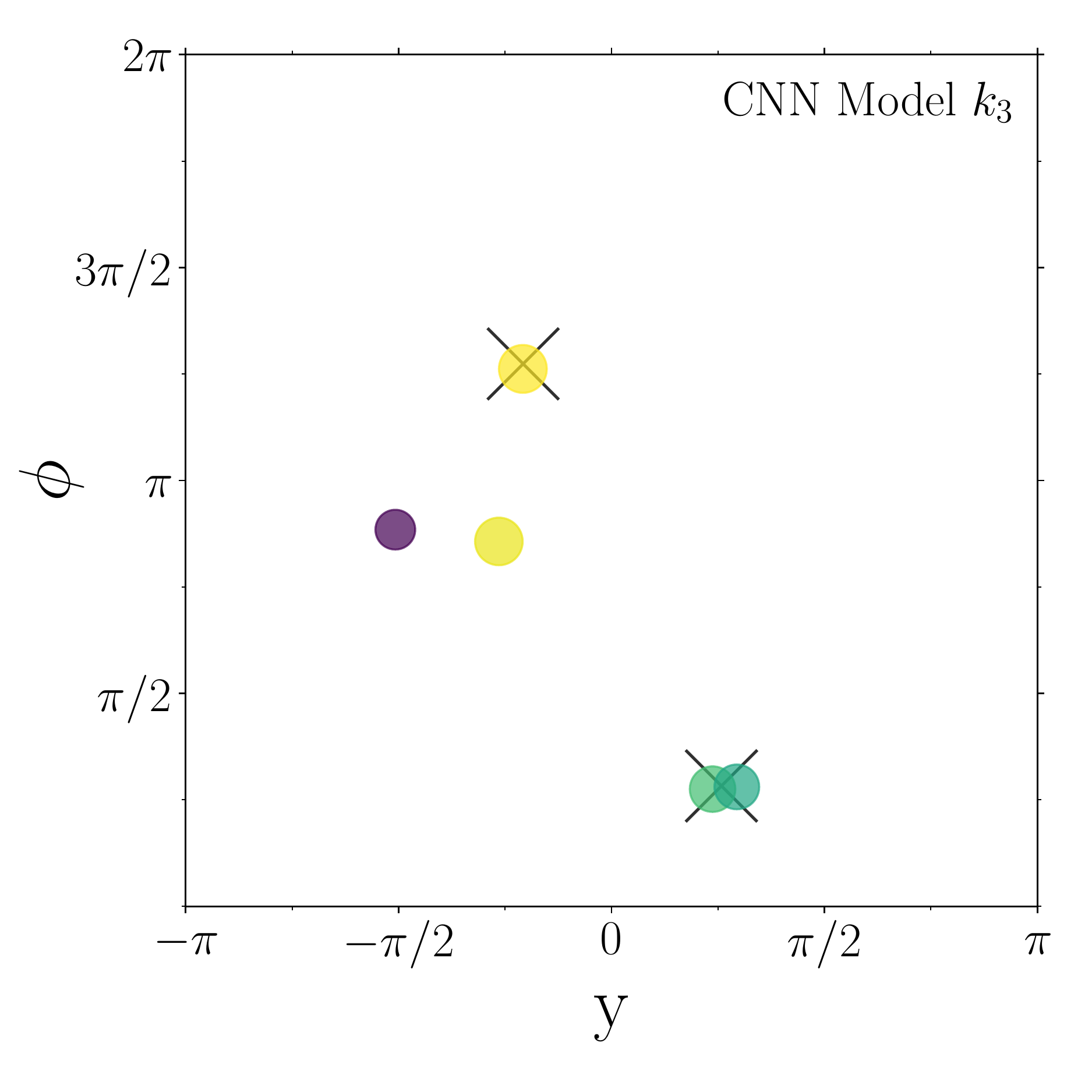}\\
\includegraphics[width=0.42\textwidth]{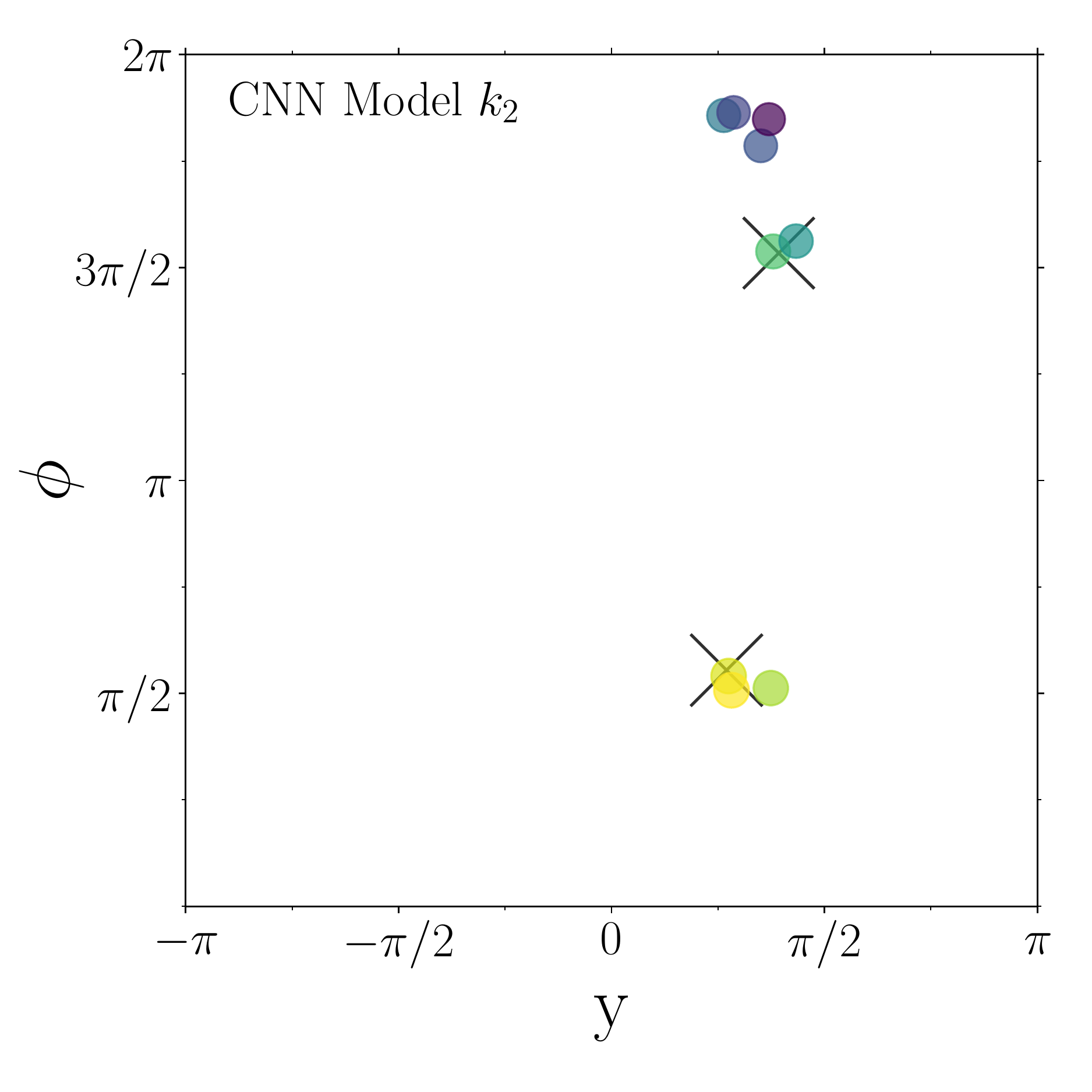}
\includegraphics[width=0.42\textwidth]{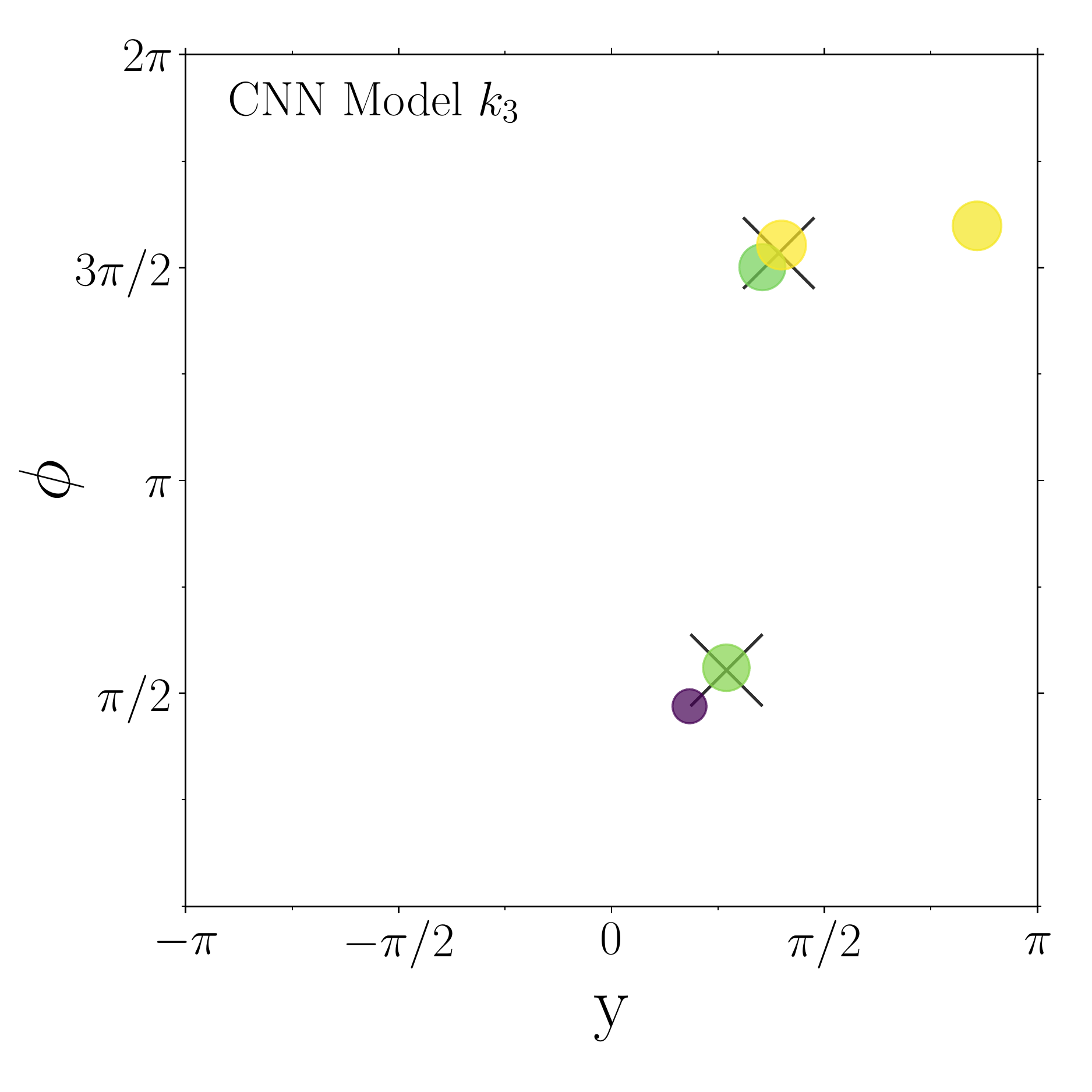}
\caption[CNN event displays]{Three example emission patterns generated by the $k_{2}$ (left column) and $k_{3}$ (right column) CNN models.  The coloured circles indicate the location of emitted particles, and the sizes and colours of the circles indicate the particle \pT.  The grey Xs in each plot are the matrix element emissions that are input to the CNN.}\label{fig:eventdisplays}
\end{center}
\end{figure}

Some example emission patterns  generated by the CNN models $k_{2}$ and $k_{3}$ are shown in Figure \ref{fig:eventdisplays} for events that satisfy the anti-$k_{t}$ R=0.4 jet selection.  The ME partons that are used to seed the CNN are shown as grey crosses.  The CNN splits these partons into a shower of lower energy partons in the region of the initial parton.  The CNN is also able to generate some wider-angle activity; for example, in the middle panel, model $k_{2}$ has generated a jet around $\left\{y, \phi\right\} = \left\{-\pi/2, \pi/4\right\}$.  

\begin{figure}[!t]
\includegraphics[width=0.5\textwidth]{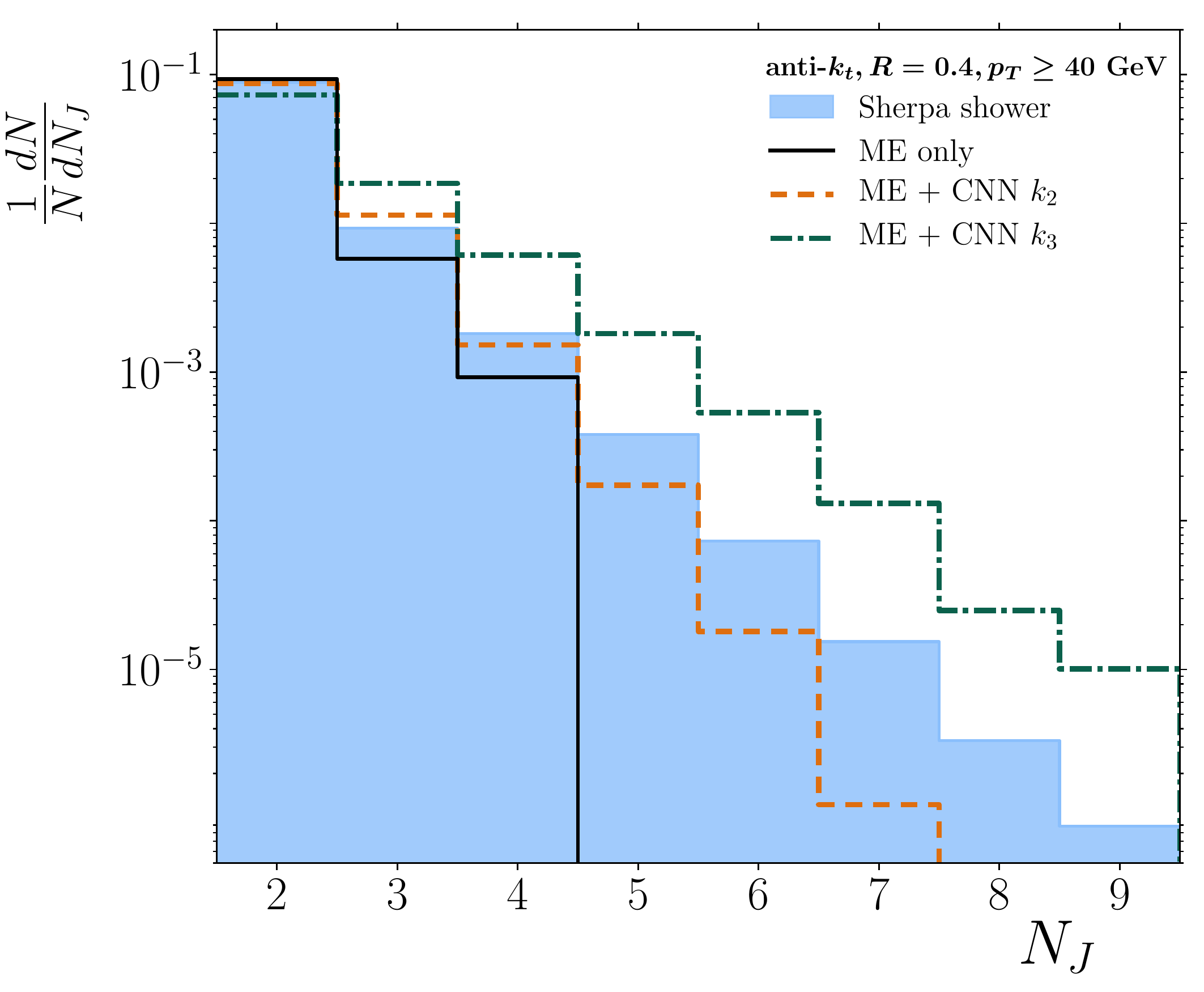}
\includegraphics[width=0.5\textwidth]{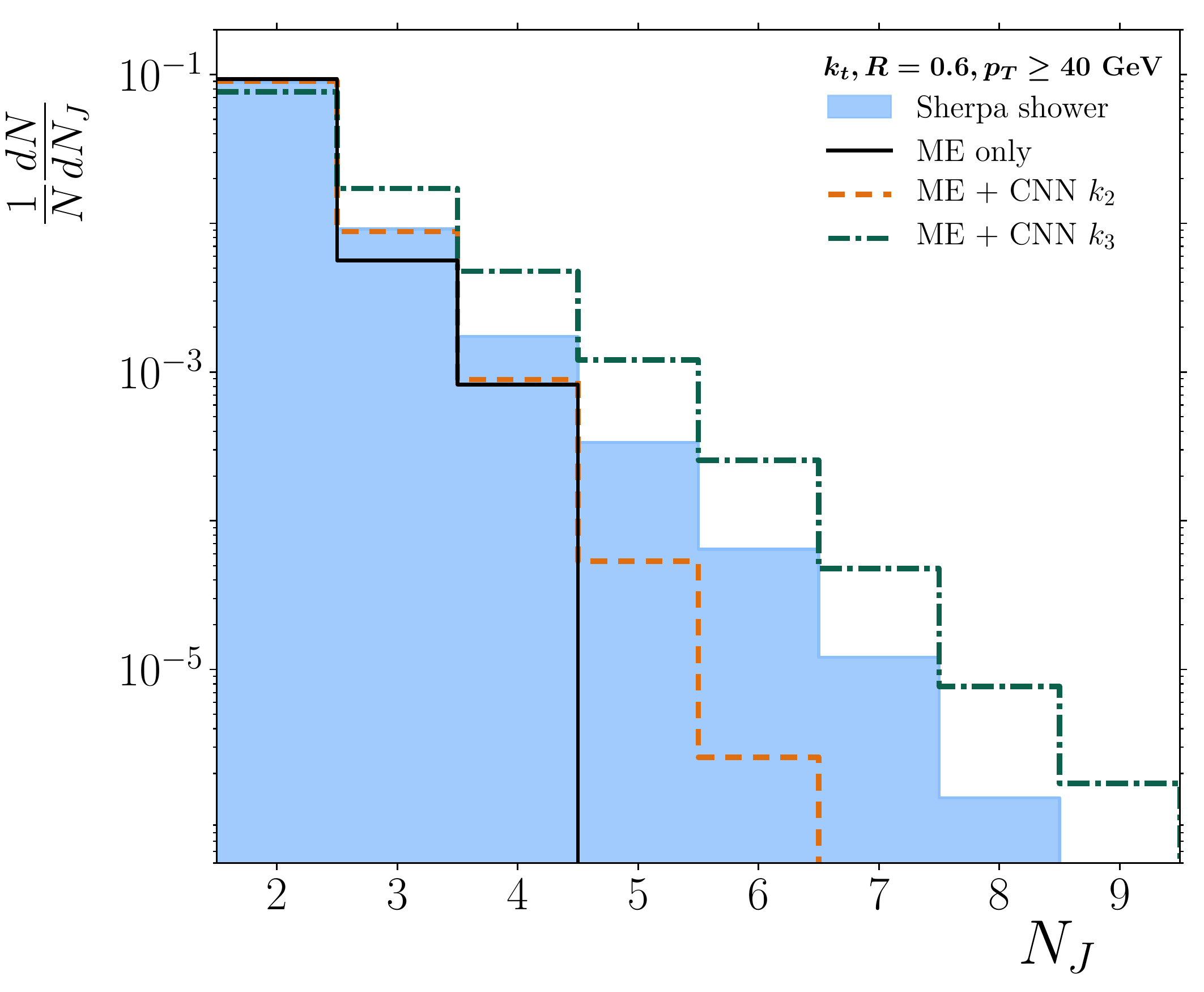}
\caption[Jet Multiplicity]{Number of jets per event using the \antikt R=0.4 jet algorithm (left) and the \kt R=0.6 algorithm (right).}\label{fig:jetmultiplicity}
\end{figure}

The distribution of the number of jets in each event that satisfy the jet requirements are shown in Figure \ref{fig:jetmultiplicity}.  The matrix element can (rarely) generate at most four partons, so events that contain more than four jets have had those jets generated by the parton shower.   Model $k_{2}$ produces somewhat fewer high multiplicity events than the target Sherpa model, while model $k_{3}$ produces slightly too many high multiplicity events.  It is encouraging that the two CNN models bracket the target data, which suggests that a CNN could indeed be made to describe the jet multiplicity after further adjustment and improvement.

\begin{figure}[!ht]
\includegraphics[width=0.5\textwidth]{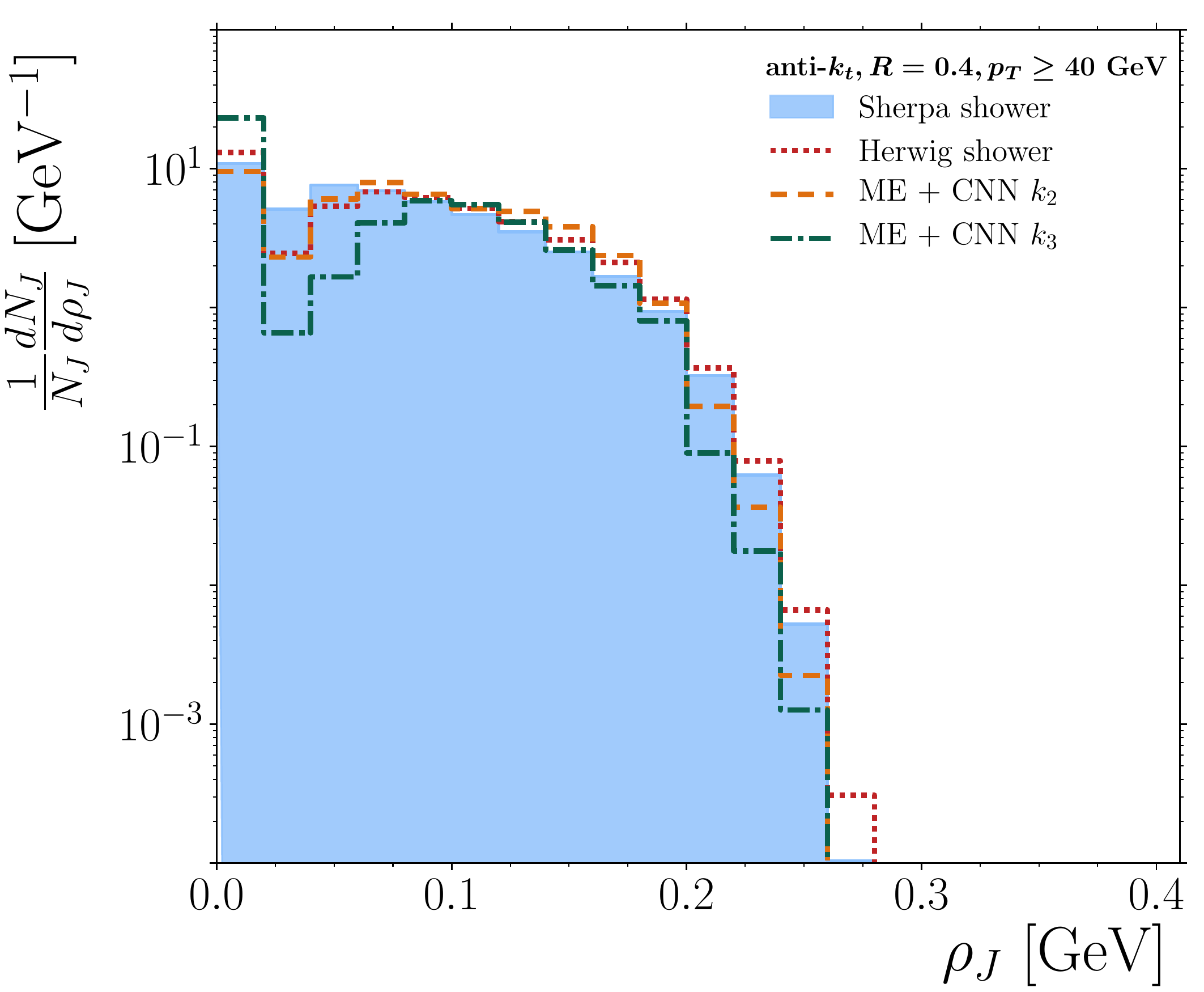}
\includegraphics[width=0.5\textwidth]{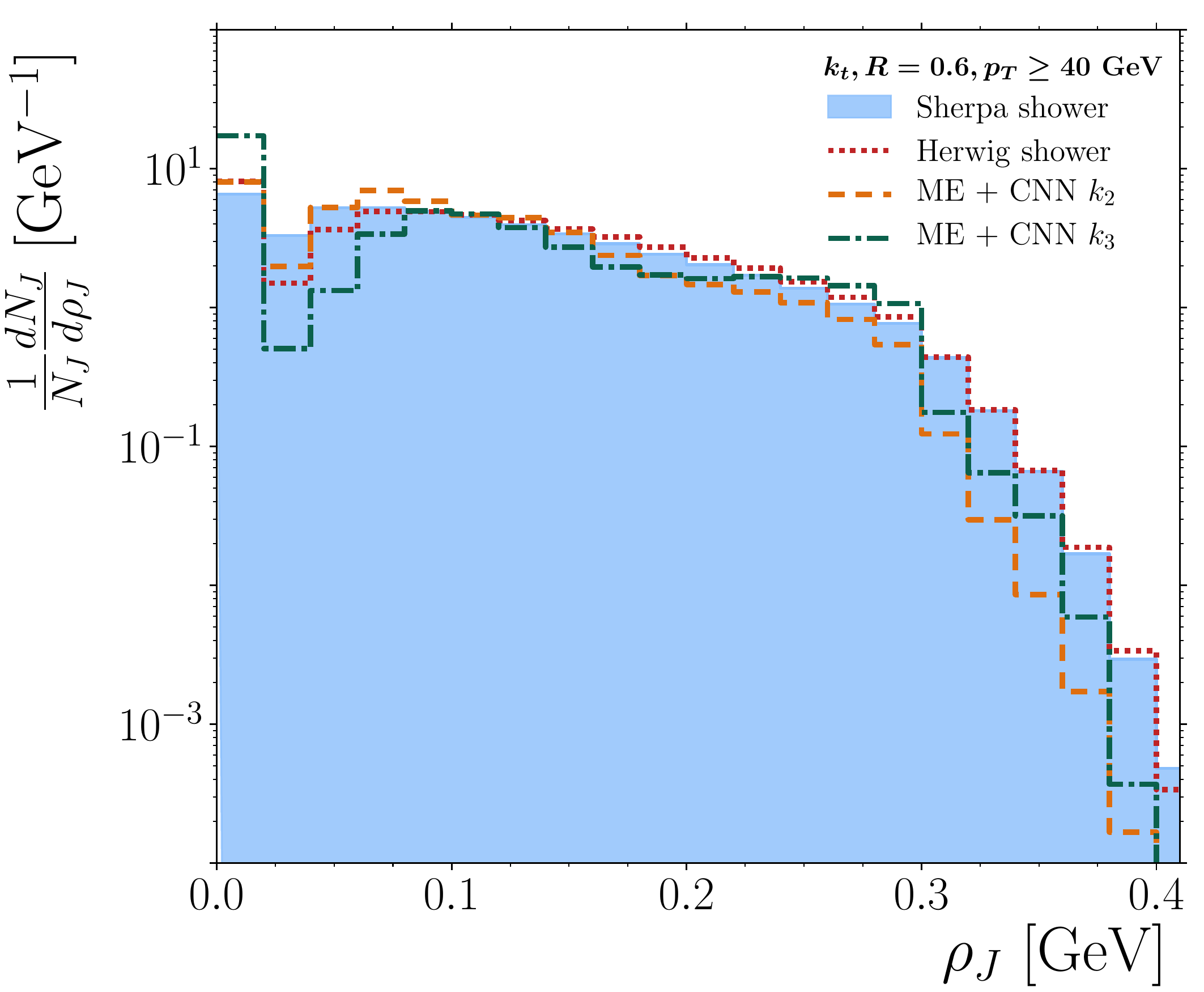}
\caption[Inclusive jet width]{Jet width distributions using the \antikt R=0.4 jet algorithm (left) and the \kt R=0.6 algorithm (right).}\label{fig:jetwidth_incl}
\end{figure}

The jet width,\footnote{$\rho$ is given by $\rho = \frac{\sum\limits_{i}\Delta R\left(j, p^{i}\right)p_{T}^{i}}{\sum\limits_{i} p_{T}^{i}}$ where the sum is over all constituents of the jet, $p_{T}^{i}$ is the \pT of the $i^{\mathrm{th}}$ jet constituent and $\Delta R\left(j, p_{i}\right)$ is the angular separation between that constituent and the jet axis. }  $\rho$, is a test of the shape of the radiation pattern emitted around a jet, and is shown in Figure \ref{fig:jetwidth_incl} for all jets that satisfy the selection criteria.  The simple CNN models do a surprisingly good job of recreating the jet shapes of the true parton shower, especially for the large width jets.  There is a deficiency in small width jets compared to Sherpa, and an over-abundance of zero-width jets.  This suggests some kind of dead cone effect, which could be an artefact of the approximate merging procedure, or some other effect of using an angular ordered-type shower.  By way of comparison, Herwig's angular-ordered shower also displays a similar dip in the number of low width jets and shows the range of expected differences between an angular-ordered shower and a $k_{T}$ ordered shower.   The CNN models have no information about parton mass, and also have a cut off at small angle due to the finite pixel size, both of which may affect the small width jets to some extent.

\begin{figure}[!ht]
\includegraphics[width=0.5\textwidth]{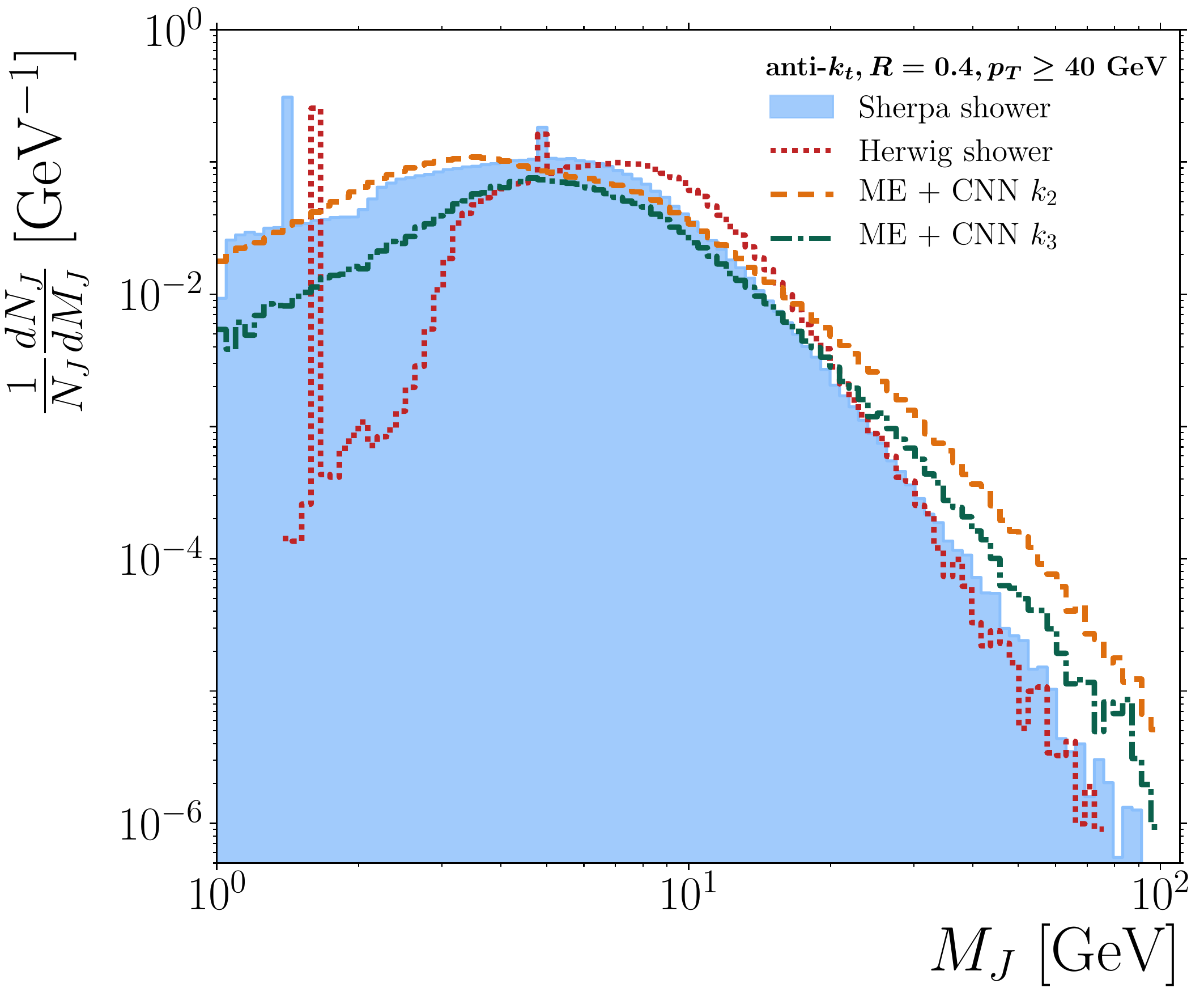}
\includegraphics[width=0.5\textwidth]{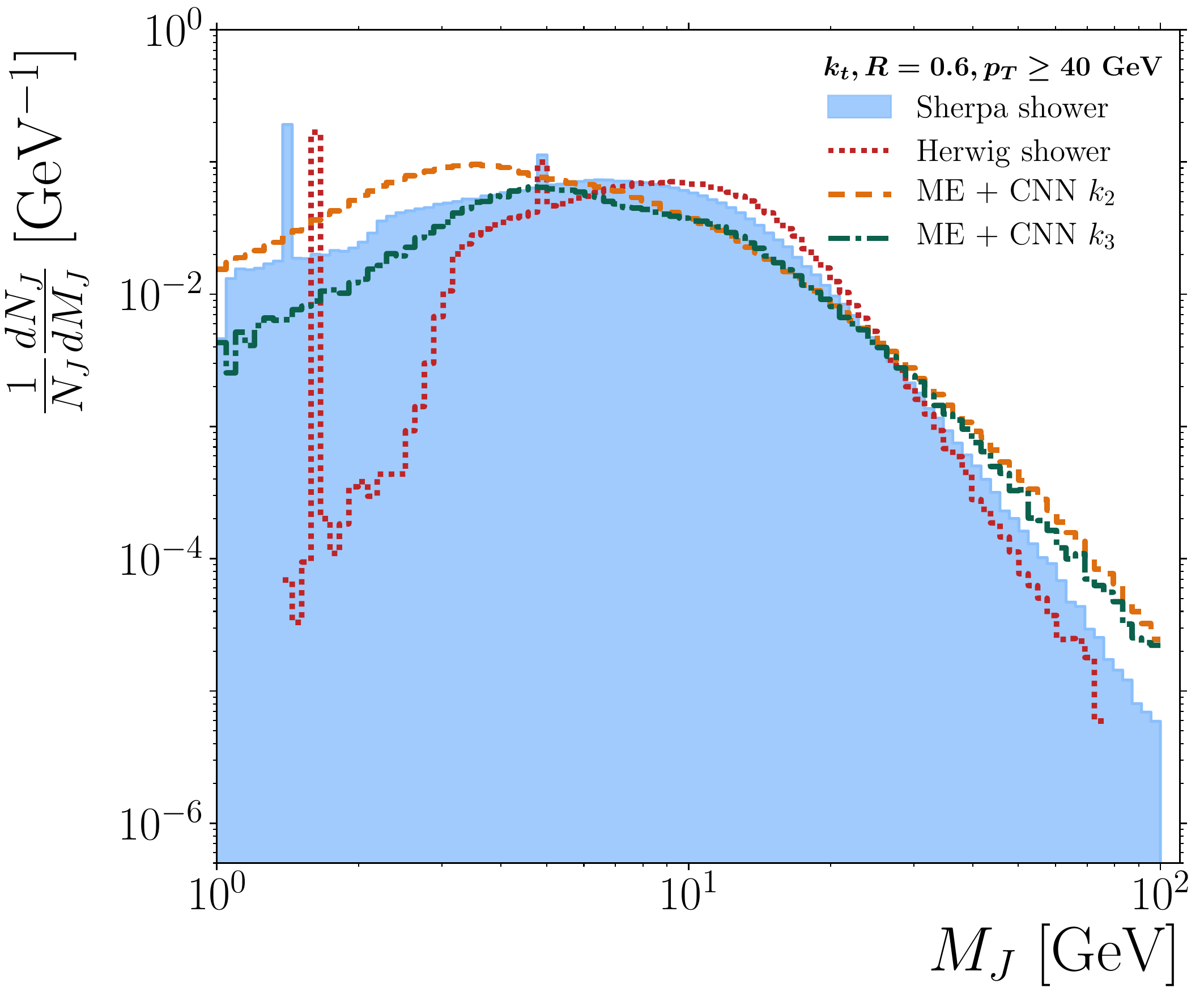}
\caption[Inclusive jet mass]{Jet mass distributions using the \antikt R=0.4 jet algorithm (left) and the \kt R=0.6 algorithm (right).}\label{fig:jetmass_incl}
\end{figure}

Jet masses arise from the finite width of the jet, and jet mass distributions also serve as a test of the radiation emitted around a jet.  The distributions of jet masses from all selected jets are shown in figure \ref{fig:jetmass_incl}.  Both the $k_{2}$ and $k_{3}$ CNN models have generated smooth mass distributions from the input ME partons, with gradients close to those of the target Sherpa model in the tails.  However, the peak of the mass distributions do not match the target.  This is not surprising because the CNN models do not contain any information about mass and do not trace the parton masses through the network; jet masses arise only from the angular width of the jets. Furthermore, the existence of massive $b$ and $c$ quarks can be seen in the Sherpa mass distribution as the small spikes at around 4.5 and 1.7~GeV, respectively.  Since the CNN does not include any mass term for the partons (or pixels) it cannot reproduce these spikes.  Again, the Herwig shower is shown as an example of the differences that can be expected between angular and $k_{T}$ ordered showers, in particular in the low mass region.

\begin{figure}[!t]
\includegraphics[width=0.5\textwidth]{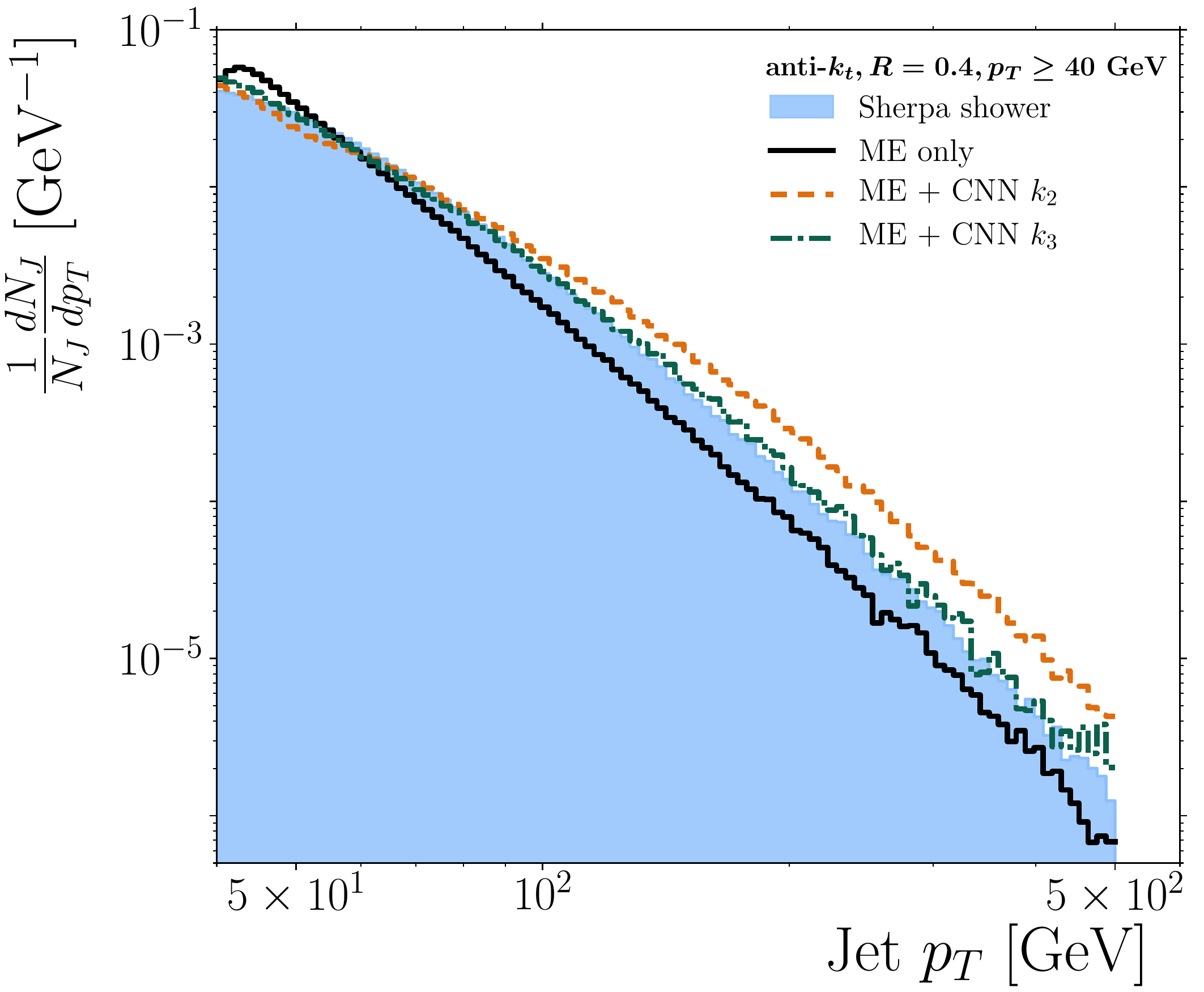}
\includegraphics[width=0.5\textwidth]{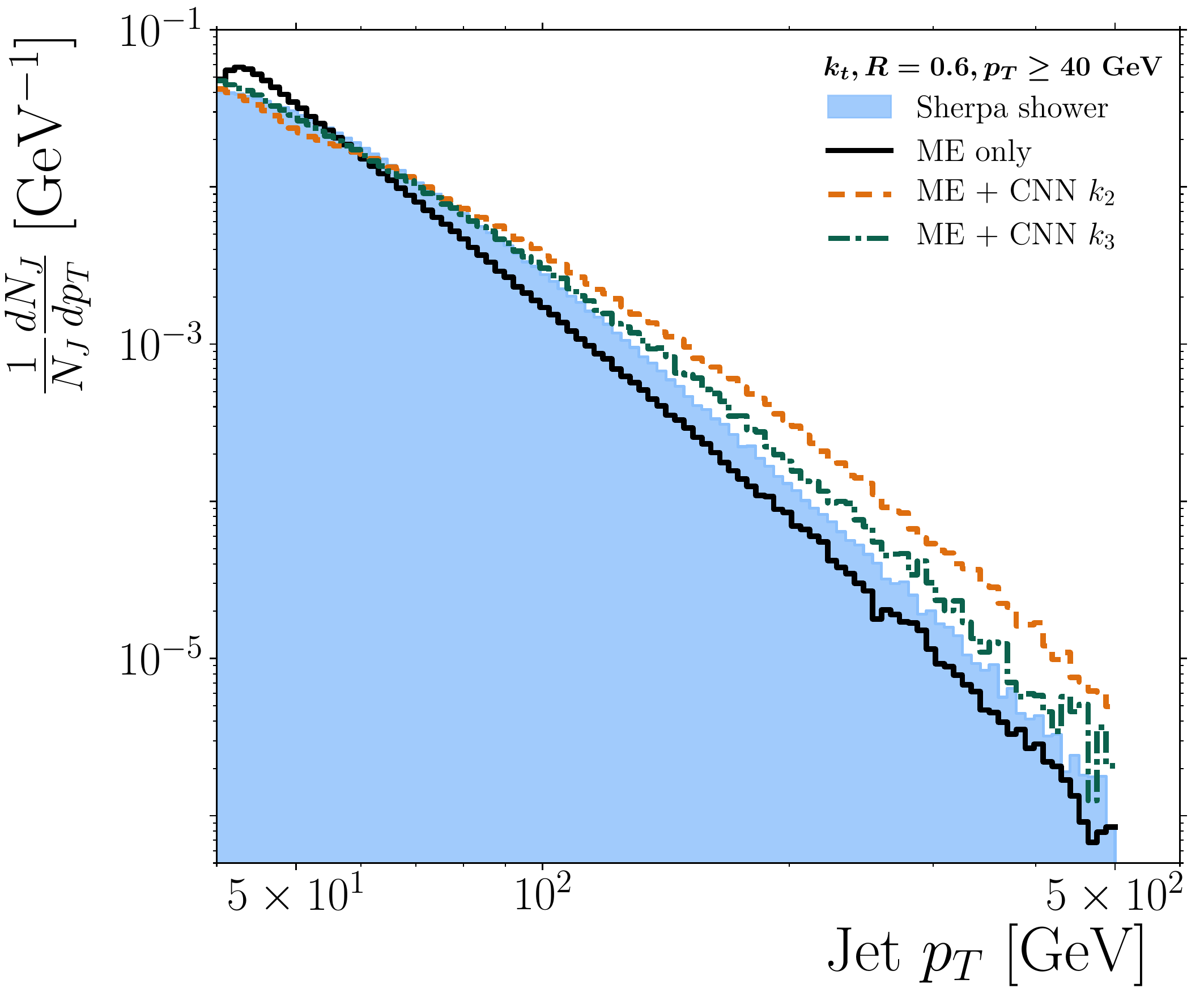}
\caption[Inclusive jet \pT]{Jet \pT distributions using the \antikt R=0.4 jet algorithm (left) and the \kt R=0.6 algorithm (right).}\label{fig:jetpt_incl}
\end{figure}

Finally, the transverse momentum (\pT) distributions of all jets that satisfy the selection criteria are shown in Figure \ref{fig:jetpt_incl}.  Both of the CNN models improve the jet \pT spectra relative to the unshowered matrix element partons by increasing the proportion of high-\pT jets and flattening the bump\footnote{This small bump occurs because the ME event selection requires the sub-leading jet to pass the same $p_{T} > 40$~GeV criterion as the leading jet.} in the ME distribution between 40 and 50~GeV.  Model $k_{3}$  is very close to the \pT spectrum of the target Sherpa parton shower for both jet algorithms.  However, model $k_{2}$ is somewhat too hard, and shows a flattening of the spectrum around 80~GeV.  This flattening is an artefact of the shower merging procedure and disappears if the merging layer is removed from the CNN.

\begin{figure}[!t]
\includegraphics[width=0.5\textwidth]{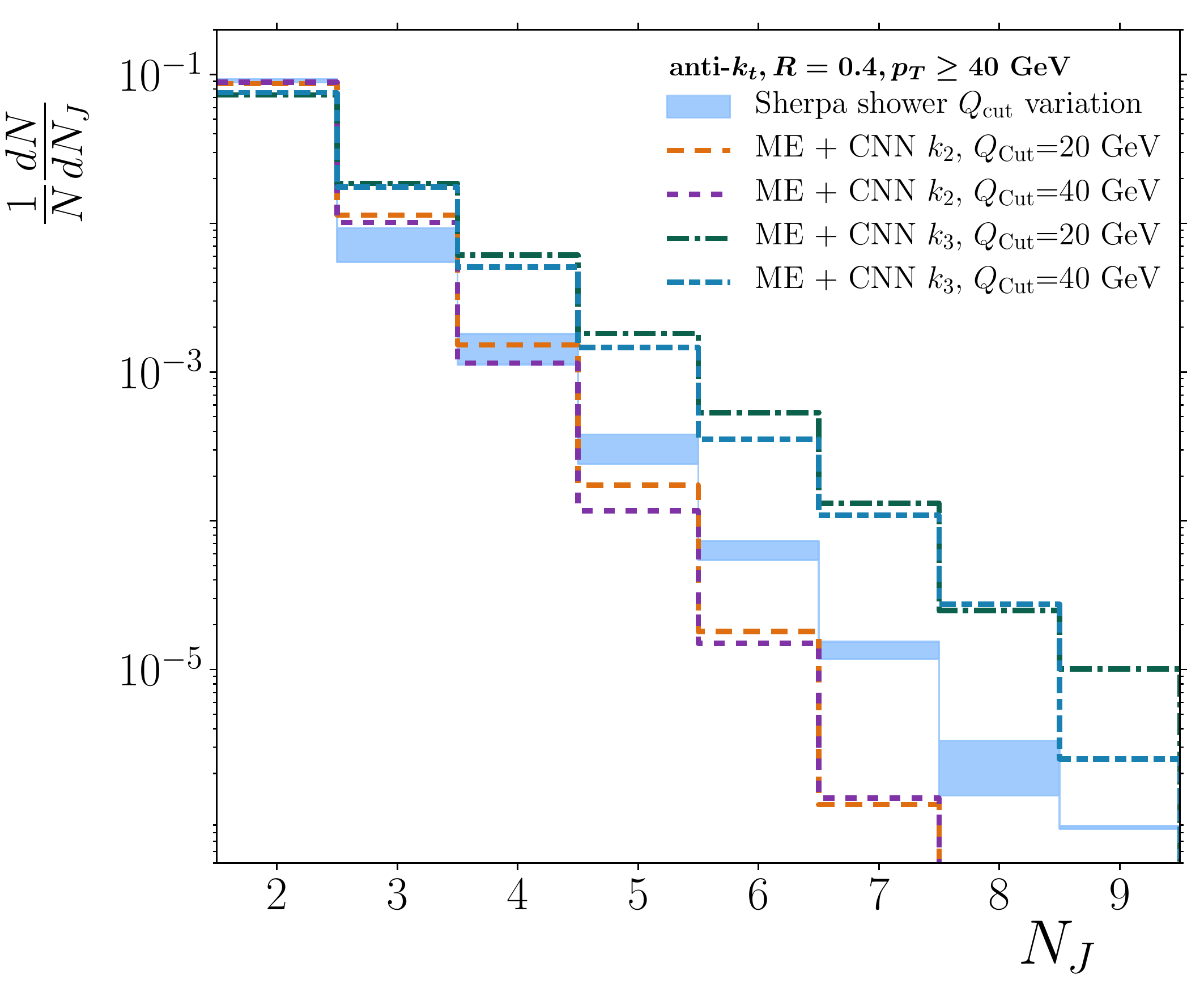}
\includegraphics[width=0.5\textwidth]{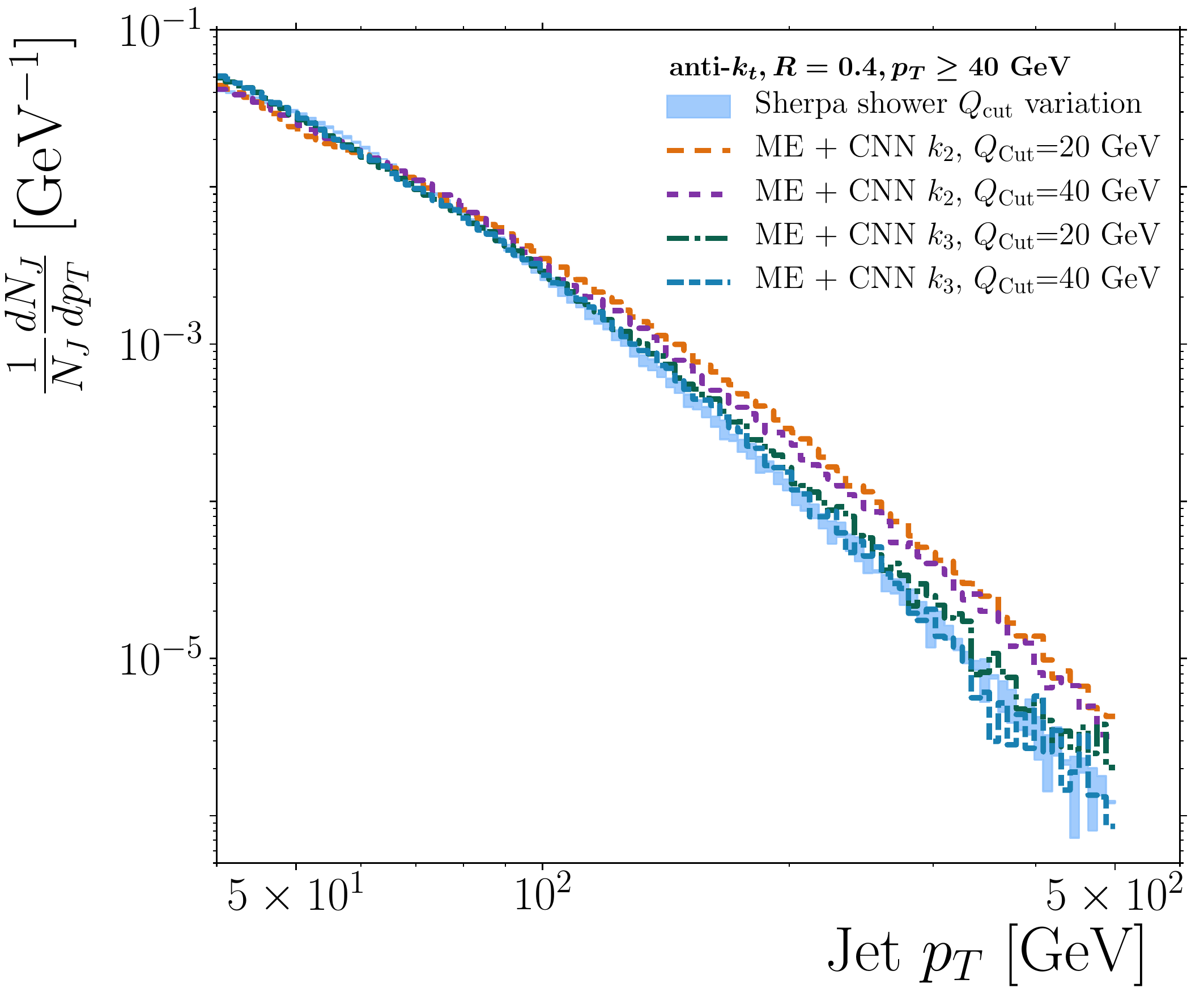}
\caption[Comparison of Merging Scales]{Jet multiplicity (left) and \pT (right) distributions compared for merging scales of 20~GeV and 40~GeV for models $k_{2}$ and $k_{3}$.  The merging scale variation using Sherpa's native shower with merging scales of 20 and 40 GeV is shown as the shaded blue band.}\label{fig:mergingScale}
\end{figure}

As a test of the shower merging procedure described in section \ref{sec:merging}, distributions for the number of jets and the jet \pT using a second set of matrix element events with a higher merging scale of $Q_{\mathrm{cut}}=40$~GeV are produced.  The shower veto scale in the network merging layer  is also updated from 20~GeV to 40~GeV to match the input matrix element calculation, and the two CNN models with this updated scale are used to shower the matrix elements.  A comparison between the results from the $Q_{\mathrm{cut}}=20$~GeV and  the $Q_{\mathrm{cut}}=40$~GeV samples for the jet multiplicity and jet \pT distributions are shown in Figure \ref{fig:mergingScale}.  While there is a small difference between the results with the two merging scales, this is consistent with the difference when using the two different merging scales with Sherpa's native shower, as shown by the shaded blue band in Figure \ref{fig:mergingScale}.  That the consistency of the CNN merging scheme at different merging scales is comparable to that of Sherpa's own shower is somewhat surprising given that the  discrete angular scales used by the CNN imply an inherent ambiguity in the evolution of the merging veto scale with network depth.

\section{Features learned by the CNN}

Each level of decomposition within the model network corresponds to a different angular scale for emissions.  Deeper layers - those closer to the compression bottleneck - capture larger angles, while layers at the very top and bottom of the network have a smaller receptive field and reproduce small angle effects.  Each level of decomposition has its own \filtermask layer, with each \filtermask storing a different set of probabilities for the filter activations at the different network depths.  Thus by converting the decomposition level - given by the network depth - to an angular scale, it is possible to study the evolution of the filter activation rates with angular scale.  The angular scale, $\Delta\phi$, to which the $l^{\mathrm{th}}$ level of the decomposition of a $N\times N$ image corresponds to is given by equation \ref{eqn:angular}

\begin{equation}
\Delta\phi = \pi \frac{k^{l}}{N}\label{eqn:angular}
\end{equation}

\begin{figure}[!t]
\includegraphics[width=0.5\textwidth]{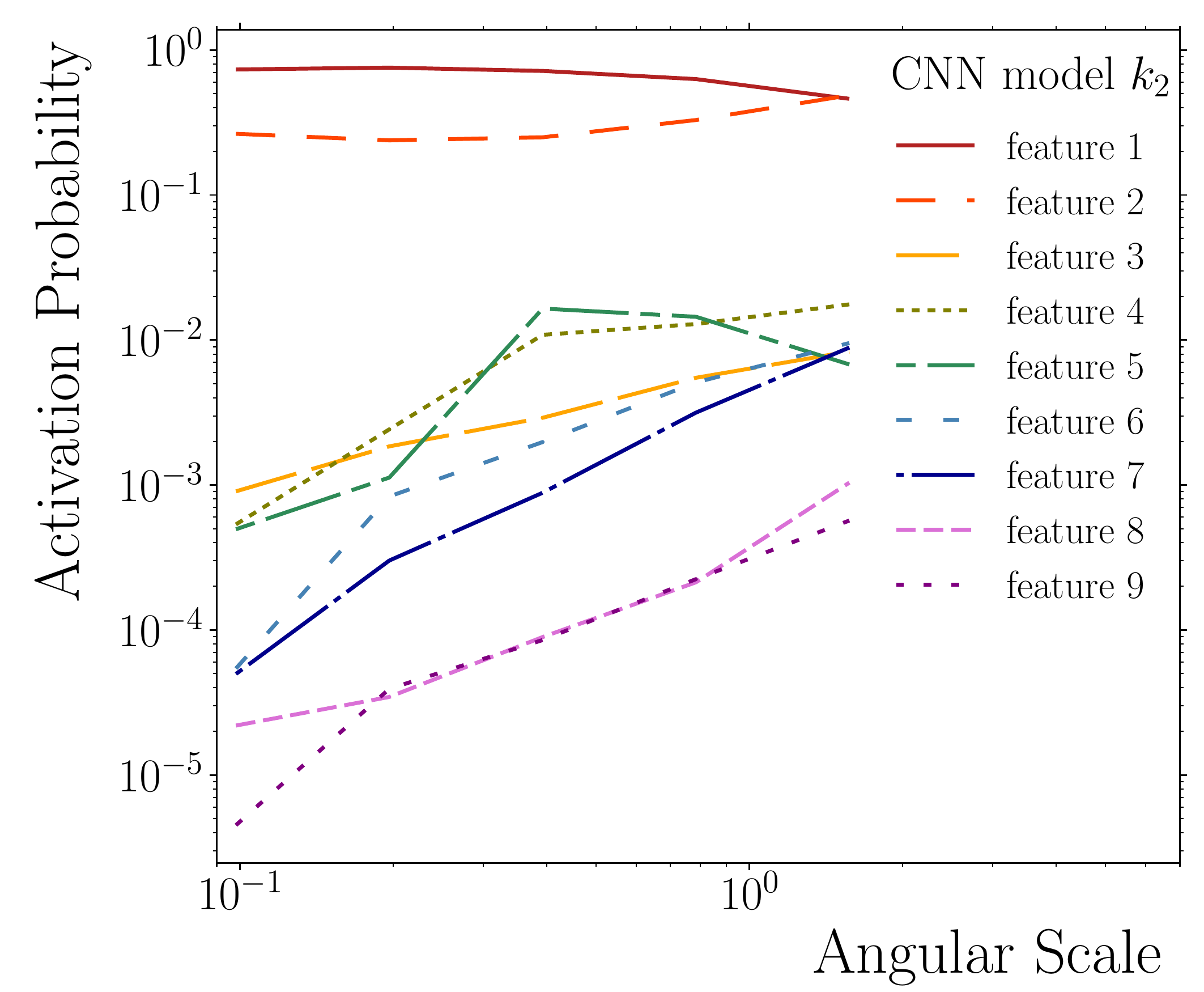}
\includegraphics[width=0.5\textwidth]{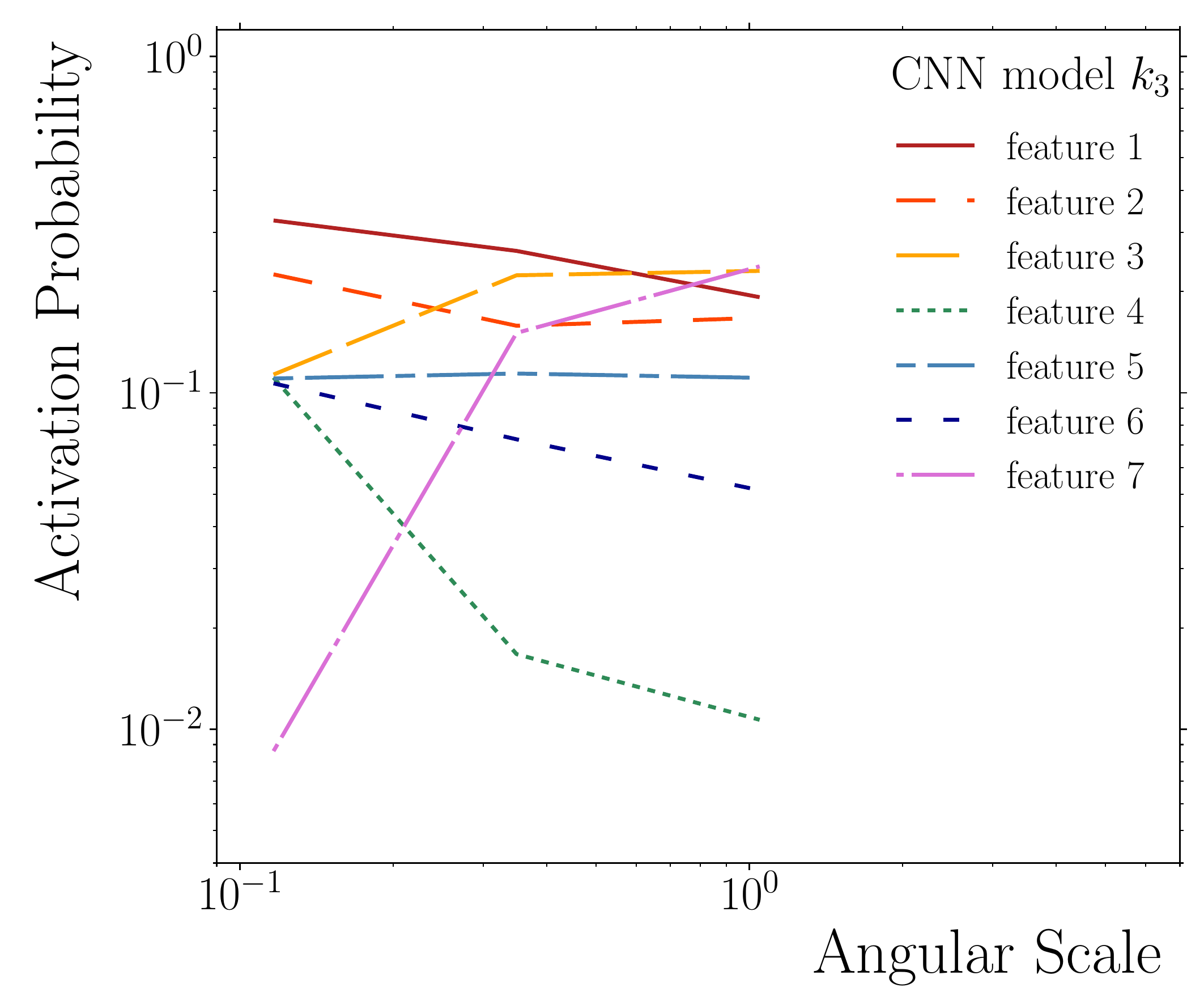}
\caption[Feature Evolution]{Evolution of the feature activation rates with angular scale for model $k_{2}$ (left) and model $k_{3}$ (right).}\label{fig:evolution}
\end{figure}

Figure \ref{fig:evolution} shows the evolution of the filter activation rates with the angular scale.  The set of $F$ individual filters  are labelled feature 1-$F$, where $F$ is the number of filters in the model and feature 1 is defined as the filter with the largest activation probability at small angles, while feature $F$ is the filter with the smallest activation probability at small angle.  Recall that the same set of filters are used at all angular scales, it is only the filter activation probability that changes with angular scale.  The activation probabilities exhibit some interesting behaviour that is suggestive of interactions between the different filters.  For example, model $k_{2}$ appears to show bands of filters having similar activation rates, while model $k_{3}$ also shows apparent correlations in the activation rates at different scales.

The features that each filter in the model encode are revealed by fixing the model weights so that only a single pair of \conv and \convtranspose filters is active.  A randomly generated pixel array is input to this  sub-model and then updated using the output of that model.  The update is repeated twenty times until a stable pixel array is converged upon.  This iterative procedure is itself repeated using one hundred different random starting arrays, and the average is taken of the results.   

\begin{figure}[!ht]
\includegraphics[width=\textwidth]{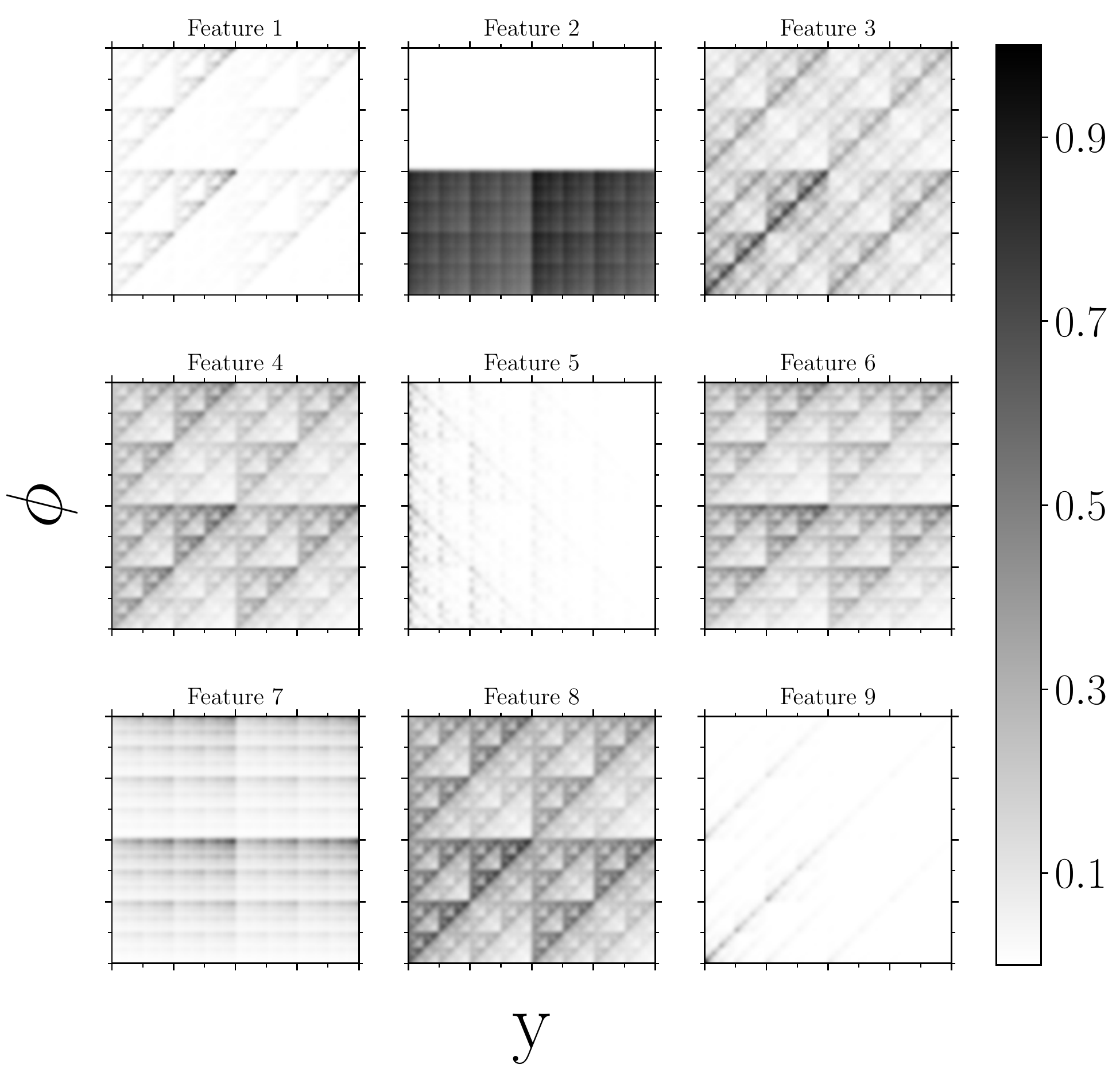}
\caption[Feature of $k_{2}$]{Features encoded in model $k_{2}$.}\label{fig:k2_features}
\end{figure}

\begin{figure}[!ht]
\begin{center}
\includegraphics[width=\textwidth]{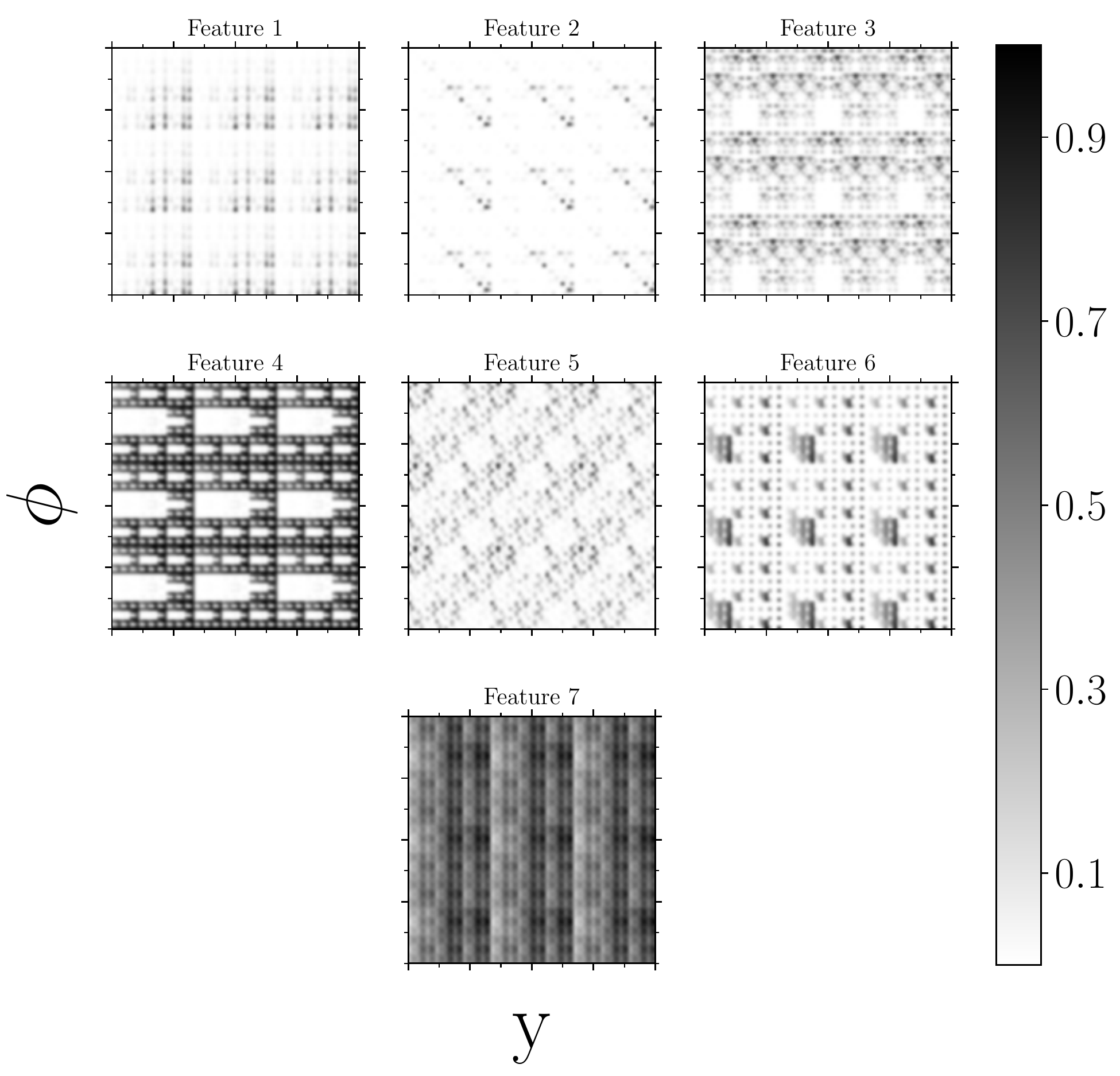}
\caption[Feature of $k_{3}$]{Features encoded in model $k_{3}$.}\label{fig:k3_features}
\end{center}
\end{figure}

The features encoded by the nine filters in model $k_{2}$ are shown in Figure \ref{fig:k2_features}, and the seven features encoded by the filters of model $k_{3}$ are shown in Figure \ref{fig:k3_features}.  All the features exhibit self-similarity, which is a result of the use of the same filter at multiple angular scales.  The features encoded in model $k_{3}$ are more complex than those in model $k_{2}$ due to the larger kernel size of the former.

\newpage
\section{Concluding Remarks}\label{sec:conclusion}

We have demonstrated that it is possible to encapsulate many of the features of a QCD parton shower in an autoencoding recursive convolutional neural-network, and that the number of trainable network parameters needed to do so is not large.   The network design is inspired by self-similarity and wavelet decomposition, which significantly reduces the number of network parameters.

The CNN learns features from jet events without needing any human supervision to classify the events or objects within them.  Convolutional neural networks have often been used as a tool for event or jet classification \cite{deOliveira:2015xxd, Komiske:2016rsd}, but here we have shown that they may also be used to learn features from QCD directly, with minimal assumptions about jet behaviour and no requirement that jets be classified using human-defined labels.  Using concepts like quark/gluon tagging as a target for neural networks in supervised learning  imposes human \mbox{(mis-)} concepts and biases onto the data, but  the basis set of features learned during the course of training a jet tagger is in many ways more interesting than the tagger itself.  Here we have attempted to ask the more general question ``what features may be learned from QCD while imposing as limited a set of biases as possible?''  

The generative model described in \cite{deOliveira:2017pjk} has a similar goal to our present work of using CNNs to learn features from jets.  However, the adversarial approach used there is quite different to our autoencoding model.  Among other things the adversarial model lacks the explicit recursion that limits the number of model parameters.  Recursively using convolutional kernels as analogues of shower splitting functions and interpreting depth within the network as the splitting angle allows the autoencoding CNN to be merged with a fixed-order matrix element to shower entire collision events.  On the other hand, the adversarial model can only generate single jet images.  There may even be the potential to combine the two approaches in the same way that traditional event generators combine separate models for showering, MPI and hadronisation.  The output of an autoencoding shower model could be passed to a further generative model to add non-perturbative details that are missing from the simpler - but more easily interpreted - recursive shower picture.

Recursive (but not convolutional) networks that use particle four-vectors as inputs have also been used both to classify jets \cite{Louppe:2017ipp} and to learn particle probability distributions that can form the basis of a generative model in JUNIPR \cite{Andreassen:2018apy}.  The JUNIPR model takes its hierarchical ordering from a jet clustering algorithm, and although that model has enough flexibility to work with any clustering hierarchy, the performance does show some residual dependance on the choice of jet algorithm.  In contrast, an angular (but not jet-derived) hierarchy is built into the structure of our autoencoding model.  JUNIPR and our model use similar ideas to capture features of QCD, though expressed in slightly different ways with slightly different aims.  Having shown that recursion can be directly included in image-based models, it may be that vector-based and image-based models evolve towards the same sort of model structure.  Indeed, recent work has shown that convolutional models may soon be used for tasks for which recurrent or recursive models have traditionally been used, such as language translation \cite{2018arXiv180803867E}.

Convolutional networks using images have sometimes been criticised as a tool for QCD because they use a large number of learned network weights.  We have shown that a large number of learned CNN parameters is not necessary to describe QCD, and that self-similarity can be implemented with recursion in image-based models. 

This method also bears some comparison with shower deconstruction \cite{Soper:2012pb}.  In shower deconstruction, a simplified parton shower model is used to estimate the probability that a given parton configuration originates from either signal or background.  The probability is estimated by evaluating the history of the splitting terms that lead to the final parton configuration.  Similarly, the compression stage of the autoencoder matches a series of learned kernels to the input parton configuration, taking the best match at each stage via the max-pooling operation.  In this way, the autoencoder tests, in parallel, a very large number of possible shower histories.  Similarly to shower deconstruction, an autoencoder trained on QCD background data could be used to discriminate between background and signal because the non-QCD signal will not match the learned shower behaviour.

Though far from perfect, the simplified models used here are able to capture qualitatively the behaviour of the Sherpa target on which they are trained.   The models could equally have been trained directly on appropriate jet data from the Large Hadron Collider, with the proviso that some important non-perturbative effects (hadronisation and MPI) were (intentionally) left out here.  This raises the interesting possibility that, with further improvement - particularly by adding a mass term - deep neural networks can provide a new way of learning about QCD that allow for data-driven background models that do not use any assumptions beyond the basic network structure.

\acknowledgments

Thanks to Alejandro Alonso for helping to configure a pair of GPUs for TensorFlow, and for technical suggestions related to Keras.  Thanks also to Frank Krauss for a useful discussion and suggestions about shower merging and shower deconstruction. Finally, thanks to Peter Hansen, Troels Petersen and Stefania Xella, who provided useful feedback that improved the paper.  This work was funded by the Danish National Research Foundation.

\bibliography{shower}
\bibliographystyle{JHEP}

\end{document}